\documentclass[journal]{IEEEtran}

\usepackage{amsmath,amssymb,graphicx,enumitem,mathtools,subcaption,comment,multirow,blkarray,booktabs,bm,stfloats,xcolor}
\usepackage{amsthm,enumitem,array}
\usepackage{longtable}
\usepackage{url}
\usepackage{breakurl}
\usepackage[colorlinks=true,citecolor=blue,linkcolor=blue,urlcolor=blue,bookmarks,bookmarksopen,bookmarksdepth=2,backref=page,breaklinks]{hyperref}

\DeclareCaptionSubType * [alph]{table}
\usepackage{tabularx}
\usepackage{lipsum}
\newcolumntype{L}[1]{>{\raggedright\arraybackslash}p{#1}}
\newcolumntype{C}[1]{>{\centering\arraybackslash}p{#1}}
\newcolumntype{R}[1]{>{\raggedleft\arraybackslash}p{#1}}

\usepackage[labelsep=period]{caption}

\usepackage{algpseudocode}
\usepackage[Algorithmus]{algorithm}
\algnewcommand{\IfOneRow}[1]{\State\algorithmicif\ #1,}
\algnewcommand{\EndifOneRow}{}

\algdef{SE}[SUBALG]{Indent}{EndIndent}{}{\algorithmicend\ }%
\algtext*{Indent}
\algtext*{EndIndent}

\makeatletter
\renewcommand{\ALG@name}{Algorithm}
\makeatother

\let\OLDthebibliography\thebibliography
\renewcommand\thebibliography[1]{
	\OLDthebibliography{#1}
	\setlength{\parskip}{0pt}
	\setlength{\itemsep}{0pt plus 0.3ex}
}

\setlist{leftmargin=*, itemsep=2pt, topsep=2pt, parsep=0pt, partopsep=5pt}

\newcommand*{\doi}[1]{\href{\detokenize{#1}}{doi: \detokenize{#1}}}
\renewcommand*{\backref}[1]{}
\renewcommand*{\backrefalt}[4]{%
	\ifcase #1 (Not cited.)%
	\or        (Cited on page~#2.)%
	\else      (Cited on pages~#2.)%
	\fi}

\DeclareMathOperator{\snf}{SNF}
\DeclareMathOperator{\ext}{Ext}

\DeclareMathOperator{\dev}{dev}

\DeclareMathOperator{\Aut}{Aut}

\DeclareMathOperator{\wt}{wt}
\DeclareMathOperator{\rank}{rank}

\DeclareMathOperator{\tworank}{2-rank}
\DeclareMathOperator{\Grank}{\Gamma-rank}
\DeclareMathOperator{\supp}{supp}

\theoremstyle{plain}
\newtheorem{theorem}{Theorem}[section]

\newtheorem{proposition}[theorem]{Proposition}

\theoremstyle{definition}

\newtheorem{example}[theorem]{Example}
\newtheorem{openproblem}[theorem]{Open Problem}

\newtheorem{definition}[theorem]{Definition}
\newtheorem{remark}[theorem]{Remark}

\numberwithin{theorem}{section}
\numberwithin{equation}{section}
\numberwithin{table}{section}
\numberwithin{figure}{section}

\floatname{algorithm}{Algorithm}
\numberwithin{algorithm}{section}

\newcommand{\F}{\mathbb F}
\newcommand{\Z}{\mathbb Z}

\def \II {{\bf I}}
\def \JJ {{\bf J}}

\DeclareMathOperator{\nl}{nl}

\usepackage{bookmark}
\bookmarksetup{
	numbered, 
	open,
}

\newcommand{\aBold}{\mathbf{a}}
\newcommand{\bBold}{\mathbf{b}}
\newcommand{\cBold}{\mathbf{c}}
\newcommand{\dBold}{\mathbf{d}}
\newcommand{\eBold}{\mathbf{e}}
\newcommand{\fBold}{\mathbf{f}}
\newcommand{\vBold}{\mathbf{v}}
\newcommand{\wBold}{\mathbf{w}}
\newcommand{\xBold}{\mathbf{x}}
\newcommand{\yBold}{\mathbf{y}}
\newcommand{\zBold}{\mathbf{z}}

\newcommand{\C}{\mathcal{C}}
\newcommand{\D}{\mathbb D}

\newcommand{\BB}[2]{\mathcal{B}_{{#1},{#2}}}
\newcommand{\AB}[2]{\mathcal{AB}_{{#1},{#2}}}
\newcommand{\BBnm}{\BB{n}{m}}
\newcommand{\ABnm}{\AB{n}{m}}
\newcommand{\AAnm}[2]{\mathcal{A}_{{#1},{#2}}}
\newcommand{\Affine}{\AAnm{n}{m}}

\newcommand{\graph}[1]{\mathcal{G}_{#1}}
\newcommand{\support}[1]{\mathcal{D}_{#1}}
\newcommand{\negation}[1]{\bar{#1}}

\begin{document}

\title{On design-theoretic aspects\\of Boolean and vectorial bent functions}
\author{Alexandr A. Polujan, Alexander Pott
\thanks{The authors are with the Institute of Algebra and Geometry, Faculty of Mathematics, Otto von Guericke University Magdeburg, Universit\"atsplatz 2, 39106 Magdeburg, Germany (email: alexandr.polujan@ovgu.de; alexander.pott@ovgu.de).}
}

\maketitle

\begin{abstract}
There are two construction methods of designs from {$(n,m)$}-bent functions, known as translation and addition designs. In this paper we analyze, which equivalence relation for Boolean bent functions, i.e. {$(n,1)$}-bent functions, and vectorial bent functions, i.e. {$(n,m)$}-bent functions with $2\le m\le n/2$, is coarser: extended-affine equivalence or isomorphism of associated translation and addition designs. First, we observe that similar to the Boolean bent functions, extended-affine equivalence of vectorial $(n,m)$-bent functions and isomorphism of addition designs are the same concepts for all even $n$ and $m\le n/2$. Further, we show that extended-affine inequivalent Boolean bent functions in $n$ variables, whose translation designs are isomorphic, exist for all $n\ge6$. This implies, that isomorphism of translation designs for Boolean bent functions is a coarser equivalence relation than extended-affine equivalence. However, we do not observe the same phenomenon for vectorial bent functions in a small number of variables. We classify and enumerate all vectorial bent functions in six variables and show, that in contrast to the Boolean case, one cannot exhibit isomorphic translation designs from extended-affine inequivalent vectorial $(6,m)$-bent functions with $m\in\{ 2,3 \}$.
\end{abstract}


\section{Introduction}
\label{section: 1 Introduction}
\IEEEPARstart{B}{oolean} and vectorial bent functions, also known as perfect nonlinear functions~\cite{Nyberg91,ROTHAUS1976300}, are special mappings of finite fields, which have the maximum Hamming distance from the set of all affine functions. Being optimal discrete structures, they have numerous applications in combinatorics, cryptography, coding and design theory. Particularly, the interaction between design theory and the theory of perfect nonlinear functions is of special interest. For instance, any new construction of bent functions may lead to a new construction of certain designs. On the other hand, combinatorial invariants of incidence structures constructed from functions over finite fields serve as good distinguishers between inequivalent functions and even classes of functions~\cite{EdelP09,Li20VanishingFlats,Weng20071096}. Before we briefly mention the main constructions of designs from bent functions and their most notable applications, we would like to point the reader's attention, that the notation we use below for translation and addition designs of bent functions will be introduced in details in the following sections.

A translation design of a function $F\colon\F_2^n\rightarrow\F_2^m$ (not necessarily perfect nonlinear) is defined as the development $\dev(A)$ of a certain set $A$, which is constructed from the function $F$ and has a nice combinatorial structure~\cite{DBLP:conf/ima/EdelP09,EdelP09}. The classical choice of a set $A$ for a Boolean bent function $f\colon\F_2^n\rightarrow\F_2$ is either the support $\support{f}$, a $(2^{n},2^{n-1} \pm 2^{n/2-1},2^{n-2} \pm 2^{n/2-1})$ difference set, or the graph $\graph{f}$, a $\left(2^{n}, 2, 2^{n}, 2^{n-1}\right)$ relative difference set, while for the vectorial function $F\colon\F_2^n\rightarrow\F_2^m$ one considers only the graph $\graph{F}$, which is a $\left(2^{n}, 2^{m}, 2^{n}, 2^{n-m}\right)$ relative difference set. The addition design $\D(F)$ of a perfect nonlinear function $F\colon\F_2^n\rightarrow\F_2^m$ is defined as the design, supported by codewords of the minimum weight of the first-order Reed-Muller code, appended by the function $F$, see~\cite{Bending1993,DillonSchatz1987,Ding19BCD,Kantor1985}. 
In this way, one can construct three designs from a Boolean bent function $f$ on $\F_2^n$: two translation designs $\dev(\support{f}),\dev(\graph{f})$ and one addition design $\D(f)$. However, for a vectorial bent function $F\colon\F_2^n\rightarrow\F_2^m$ there are only two options: one translation design $\dev(\graph{F})$ and one addition design $\D(F)$.

So far, translation and addition designs are used in the context of extended-affine equivalence of Boolean bent functions, which is known as the most general equivalence relation for Boolean functions. For instance, Weng et al. in~\cite[Theorem 5.11]{Weng20071096} used the 2-rank of the translation design $\dev(\support{f})$ to prove, that almost every Desarguesian partial spread bent function is not extended-affine equivalent to a Maiorana-McFarland bent function. Recently, the authors of this paper in~\cite{PolujanPott19DCC} used algebraic invariants of  $\dev(\support{f})$ and $\dev(\graph{f})$ to show inequivalence of certain homogeneous cubic bent functions. Bending in~\cite[Corollary 10.6]{Bending1993} proved that extended-affine equivalence of bent functions coincides with isomorphism of addition designs. As a useful application of this result, one can use computer algebra systems, e.g.  Magma~\cite{MR1484478} and GAP~\cite{Soicher19}, to check effectively the equivalence of bent functions in a small number of variables via the isomorphism of addition designs. Bending in~\cite[Theorems 8.4, 8.13]{Bending1993} used invariants of the addition design $\D(f)$ to derive a necessary condition for a bent function $f$ on $\F_2^n$ to be extended-affine equivalent to a Maiorana-McFarland bent function. Despite the fact, that the translation and the addition  designs of vectorial bent functions are defined in the same way as for Boolean bent functions, there are no similar applications for vectorial bent functions so far.

The main goal of this paper is to compare Boolean and vectorial bent functions from the point of view of differences between extended-affine equivalence and isomorphism of the addition and the translation designs. For instance, in the Boolean bent case isomorphism of addition designs carries all the information about the extended-affine equivalence of Boolean bent functions, and vice versa. Our first objective is to prove, that the same phenomenon occurs for addition designs of vectorial bent functions. In general, isomorphic incidence structures do not necessarily come from equivalent difference sets: Edel and Pott in~\cite[Example 1]{DBLP:conf/ima/EdelP09} observed an example of extended-affine inequivalent Boolean bent functions $f$ and $f'$ on $\F_2^6$, whose translation designs $\dev(\support{f})$ and $\dev(\support{f'})$ are isomorphic. However, these incidence structures do not have a proper generalization for the vectorial case. Our second objective is to extend the observation of Edel and Pott for translation designs $\dev(\graph{f})$ and $\dev(\graph{f'})$ of Boolean functions, i.e. find a pair of Boolean bent functions $f,f'$ on $\F_2^n$ for any $n\ge 6$, which are extended-affine inequivalent but their translation designs $\dev(\graph{F})$ and $\dev(\graph{f'})$ are isomorphic. Since the translation design $\dev(\graph{f})$ is invariant for the extended-affine equivalence, it will imply that isomorphism of translation designs $\dev(\graph{F})$ and $\dev(\graph{f'})$ of Boolean functions $f$ and $f'$ on $\F_2^n$ is a coarser equivalence relation than extended-affine equivalence. The third objective of this paper is to show, that in contrast to the Boolean case, isomorphism of designs $\dev(\graph{F})$ and $\dev(\graph{F'})$ of vectorial bent functions in six variables coincides with the extended-affine equivalence. 

After introducing the necessary background on bent functions and designs in Subsection~\ref{subsection: 1.1 Preliminaries}, we consider addition designs of vectorial bent functions. In Section~\ref{section: 2 Addition designs} we prove that similarly to Boolean bent functions, extended-affine equivalence of vectorial bent functions coincides with the isomorphism of addition designs. In this way, we solve a recent open problem, addressed by Ding, Munemasa and Tonchev in~\cite[Note 24]{Ding19BCD}. In Section~\ref{section: 3 Translation designs} we first provide examples of extended-affine inequivalent Boolean bent functions $f,f'$ on $\F_2^n$, whose translation designs $\dev(\graph{f})$ and $\dev(\graph{f'})$ are isomorphic. Consequently, we prove that for any $n\ge6$ the isomorphism of translation designs $\dev(\graph{f})$ and $\dev(\graph{f'})$ of Boolean bent functions $f$ and $f'$ on $\F_2^n$ is coarser than extended-affine equivalence. In Section~\ref{section: 4 Translation designs of vectorial bent functions} we show that the similar phenomenon does not occur for vectorial bent functions in a small number of variables. We classify and enumerate all vectorial bent functions in six variables and  observe, that in contrast to the Boolean case, vectorial bent functions $F$ and $F'$ on $\F_2^6$ are extended-affine equivalent if and only if their translation designs $\dev(\graph{F})$ and $\dev(\graph{F'})$ are isomorphic. In Section~\ref{section: 5 Conclusion and future work} we give concluding remarks and raise some open problems on bent functions and their designs. In Appendix~\ref{appendix: A} we list algebraic normal forms of the obtained representatives of equivalence classes of vectorial bent functions in six variables together with their invariants.

\subsection{Preliminaries}
\label{subsection: 1.1 Preliminaries}

Let $\F_2=\{0,1\}$ be the finite field with two elements and let $\F_2^n$ be the vector space of dimension $n$ over $\F_2$. Mappings $F\colon\F_2^n\rightarrow\F_2^m$ are called $(n,m)$-functions. The single-output case $m=1$ corresponds to \emph{Boolean} functions, while in the multi-output case $m\ge2$ one deals with \emph{vectorial} functions. The \emph{graph} $\graph{F}$ of an $(n,m)$-function $F$ is the set $\graph{F}:=\left\{(\xBold, F(\xBold)) \colon \xBold \in \F_{2}^{n}\right\}$. The \emph{support} $\support{f}$ of a Boolean function $f$ on $\F_2^n$ is the set $\support{f}:=\{\mathbf{x}\in\F_2^n\colon f(\mathbf{x})=1\}$. Any Boolean function $f\colon\F_2^n\rightarrow\F_2$ can be uniquely represented as a multivariate polynomial in the ring $\F_2[x_1,\dots,x_n]/(x_1\oplus x_1^2,\dots,x_n\oplus x_n^2)$. This representation is called the \emph{algebraic normal form} (ANF for short) and given by

$$f(\mathbf{x})=\bigoplus\limits_{\mathbf{v}\in\F_2^n}c_{\mathbf{v}} \left( \prod_{i=1}^{n} x_i^{v_i} \right),$$

\noindent where $\xBold = (x_1,\dots, x_n)\in\F_2^n$,  $c_{\vBold}\in\F_2$ and $\vBold = (v_1,\dots, v_n)\in\F_2^n$.  

There are several criteria, which an $(n,m)$-function has to satisfy in order to be considered as a good cryptographic primitive, among them are high algebraic degree and high nonlinearity. The \emph{algebraic degree} of a Boolean function $f\colon\F_2^n\rightarrow\F_2$, denoted by $\deg(f)$, is the algebraic degree of its ANF as a multivariate polynomial. This definition can essentially be extended to the vectorial case. Any vectorial function $F\colon\F_2^n\rightarrow\F_2^m$ can be uniquely (up to the choice of basis of $\F_2^m$) associated with $m$ \emph{coordinate} Boolean functions $f_i\colon\F_2^n\rightarrow\F_2$ for $1\le i\le m$ as a column-vector $F(\mathbf{x}):=(f_1(\mathbf{x}),\ldots,f_m(\mathbf{x}))^T$. In this way, the \emph{algebraic degree} of an $(n,m)$-function $F$ is defined by $\deg(F):=\max_{1\le i\le m} \deg(f_i)$. Clearly, the algebraic degree of an $(n,m)$-function $F$ can not exceed $n$.

The \emph{nonlinearity of a Boolean function} $f\colon \F_2^n\rightarrow \F_2$ is a measure of distance between the function $f$ and the set of all affine functions $\mathcal{A}_n:=\{ l\colon \F_2^n\rightarrow \F_2 | \deg(l) \le 1 \}$. Formally, it is defined as $\nl(f):=\min\limits_{l\in \mathcal{A}_n} d_H(f,l)$, where $d_H(f,g):=|\{ \xBold\in\F_2^n\colon f(\xBold)\neq g(\xBold) \}|$ is the \emph{Hamming distance} between functions $f$ and $g$ on $\F_2^n$. This definition can be extended for the vectorial case using the notion of component functions. Recall that for an $(n,m)$-function $F$, the \emph{component function} $F_{\bBold}$ is the Boolean function $F_{\bBold}\colon \F_2^n\rightarrow \F_2$, given by $F_{\bBold}(\xBold):=\langle \bBold, F(\xBold) \rangle_{m}$, where $\langle \cdot,\cdot \rangle_m$ is a non-degenerate bilinear form on $\F_2^m$. In this way, the \emph{nonlinearity of a vectorial $(n,m)$-function} $F$ is the minimum nonlinearity of all its component functions and is given by $\nl(F):=\min\limits_{l\in\mathcal{A}_n,\bBold\in\F_2^m \setminus \{ \mathbf{0}\} } d_H(F_{\bBold},l)$. The main tool to compute the nonlinearity of an $(n,m)$-function $F$ is the \emph{Walsh transform} $W_{F}\colon \F_2^n\times\F_2^m\rightarrow \Z$, defined by
$W_F(\aBold,\bBold):=\ W_{F_{\bBold}}(\aBold)$ and
\begin{equation*}
W_{F_{\bBold}}(\aBold):= \sum\limits_{\xBold\in\F_2^n} (-1)^{F_{\bBold}(\xBold)\oplus\langle \aBold, \xBold \rangle_n}
\end{equation*}
for $\aBold\in\F_2^n$ and $\bBold\in\F_2^m$. Using the Walsh transform, the nonlinearity of an $(n,m)$-function $F$ can be computed as 
$\nl(F):=2^{n-1}- \frac{1}{2}\cdot\max\limits_{\aBold\in\F_2^n,\bBold\in\F_2^m \setminus \{ \mathbf{0}\} }\left| W_F(\aBold,\bBold) \right|$.
The upper bound on nonlinearity of an $(n,m)$-function $F$ is given by $\nl( F ) \le 2 ^ { n - 1 } - 2 ^ { \frac { n } { 2 }-1}$ and functions, achieving this bound are called \emph{perfect nonlinear}.	
\begin{definition}
	An $(n,m)$-function $F$ is called \emph{bent} or \emph{perfect nonlinear} if $\nl( F ) = 2 ^ { n - 1 } - 2 ^ { \frac { n } { 2 } - 1 }$.
\end{definition}
\begin{remark}
	Throughout the paper we will call single-output bent functions, i.e. $m=1$ \emph{Boolean bent} functions, while multi-output bent functions, i.e. $m\ge2$, \emph{vectorial bent} functions. One can show that an $(n,m)$-function $F$ is bent if for all $\aBold\in\F_2^n$ and all $\bBold\in\F_2^m$ with $\bBold\neq\mathbf{0}$ the Walsh transform satisfies $W_F(\aBold,\bBold)= \pm 2^{n/2}$. Boolean bent functions exist on $\F_2^n$ if and only if $n$ is even. Vectorial $(n,m)$-bent functions exist if and only if $m\le n/2$, as it was shown by Nyberg in~\cite{Nyberg91}. The algebraic degree of an $(n,m)$-bent function is at most $n/2$, see~\cite{ROTHAUS1976300}.
\end{remark}
On the set of all $(n,m)$-functions we introduce an equivalence relation in the following way. We say that two $(n,m)$-functions $F,F'$ are \emph{extended-affine equivalent} (\emph{EA-equivalent} for short), if there exist a linear permutation $A_1$ of $\F_{2}^{m}$, an affine permutation $A_2$ of $\F_{2}^{n}$ and an affine function $A_3\colon\F_2^{n}\rightarrow\F_2^m$ such that $F=A_1\circ F' \circ A_2 \oplus A_3$. Further we will study equivalence of $(n,m)$-functions in connection with the equivalence of the associated linear codes and designs. We refer to~\cite{DesignTheoryVol1Beth} and~\cite{Ding:DesignsFromCodes} for extensive references on the subject.		

A \emph{linear code} $\C$ over $\F_2$ is a vector subspace $\C \subseteq \F_2^n$. Elements of a linear code $\mathcal{C}$ are called \emph{codewords}. The number of nonzero coordinates of a codeword $\cBold\in\mathcal{C}$ is called the \emph{weight} of $\cBold$ and is denoted by $\wt(\cBold)$. The \emph{minimum distance} of a linear code is the minimum weight of its nonzero codewords. We say, that $\mathcal{C}\subseteq \F_2^n$ is an $[n,k,d]$-\emph{linear} code, if $\mathcal{C}$ has dimension $k$ and the minimum distance $d$. The \emph{support} of a codeword $\cBold=(c_1,\ldots,c_n)\in\mathcal{C}$ is defined by $\supp(\cBold)=\left\{1 \leq i \leq n: c_{i} \neq 0\right\} \subseteq\{1,2,3, \ldots, n\}$. Two linear codes $\C$ and $\C'$ are \emph{permutation equivalent} provided there is a permutation of coordinates which sends the code $\C$ to $\C'$.

An incidence structure $\mathcal{I}=(\mathcal{P}, \mathcal{B})$ is called a $2$-$(v, k, \lambda)$ \emph{design}, if the cardinality of the point set $\mathcal{P}$ is $v$, the set of blocks $\mathcal{B}$ is a collection of \emph{$k$-subsets} of $\mathcal{P}$ and every 2-subset of points $\{ p,q \} \subset \mathcal{P}$ is contained in exactly $\lambda$ blocks of $\mathcal{B}$.  There are several generalizations of 2-designs, one of them is a divisible design. For instance, the incidence structure $\mathcal{I}=(\mathcal{P}, \mathcal{B})$ is called a $(\mu, \nu, k, \lambda)$ \emph{divisible design}, if the point set $\mathcal{P}$ with $|\mathcal{P}|=v=\mu\cdot\nu$ elements is divided into $\mu$ \emph{point classes} of size $\nu$ each, the block set $\mathcal{B}$ is a collection of $k$-\emph{subsets} of $\mathcal{P}$ and the number of blocks, containing any 2-subset $\{ p,q \}\subset \mathcal{P}$ depends on the relation between points $p$ and $q$ in the following way: if $p$ and $q$ are in the same point class, the 2-subset $\{ p, q \}$ is not contained in a block; otherwise it is contained in exactly $\lambda$ blocks. All the information about an incidence structure $\mathcal{I}$ is contained in its \emph{incidence matrix} $M(\mathcal{I})=(m_{i,j})$, which is a binary $b\times v$ matrix with $m_{i,j}=1$ if $p_j\in B_i$ and $m_{i,j}=0$ otherwise. In this way, two incidence structures $\mathcal{I}$ and $\mathcal{I}'$ are \emph{isomorphic}, if there exist permutation matrices $P$ and $Q$ such that $M(\mathcal{I})=P\cdot M(\mathcal{I}') \cdot Q$.

\section{Addition designs of Boolean and vectorial bent functions}
\label{section: 2 Addition designs}
Dillon and Schatz in~\cite{DillonSchatz1987} and Bending in~\cite[Corollary 10.6]{Bending1993} proved independently that Boolean bent functions $f$ and $f'$ on $\F_2^n$ are extended-affine equivalent if and only if their addition designs $\D(f)$ and $\D(f')$ are isomorphic. In this section we show that similar to the Boolean case, vectorial $(n,m)$-bent functions $F$ and $F'$ are extended-affine equivalent if and only if their addition designs $\D(F)$ and $\D(F')$ are isomorphic. We also use the result of Bending~\cite[Theorem 9.6]{Bending1993} to show, how one can construct an incidence matrix of the addition design of a vectorial bent function with the help of component functions and their duals. First, we give the definition of the addition design of a bent function.
\begin{definition}\label{definition: Addition designs}
	Let $F$ be an $(n,m)$-bent function and $C_{F}$ be an $(n+m+1) \times 2^{n}$-matrix  over $\mathbb{F}_{2}$, given by
	\begin{equation}\label{equation: Linear code}
	C_{F}=\left(\begin{array}{ccc}
	\cdots & 1 & \cdots \\
	\cdots & \xBold & \cdots \\
	\cdots & F(\xBold) & \cdots
	\end{array}\right)_{\xBold \in \mathbb{F}_{2}^{n}}.
	\end{equation}
	We define the linear code $\mathcal{C}(F)$ over $\mathbb{F}_{2}$, as the row space of the matrix $C_{F}$. It is not difficult to check, that the linear code $\mathcal{C}(F)$ is a $[2^n,n+m+1,2^{n-1}-2^{n/2-1}]$-code.
	Further, we define two sets $\mathcal{P}=\{ \xBold \colon \xBold \in \F_2^n \}$ and $\mathcal{B}=\{\supp(f)\colon f \in \mathcal{C}(F), \ \wt(f)=2^{n-1}-2^{n/2-1} \}$. The \emph{addition design} of an $(n,m)$-bent function $F$ is the incidence structure $\D(F)=(\mathcal{P},\mathcal{B})$, which is supported by codewords of the minimum weight of the linear code $\mathcal{C}(F)$. Ding, Munemasa and Tonchev in~\cite[Theorem 11]{Ding19BCD} proved that the addition design $\D(F)$ of an $(n,m)$-bent function $F$ is a $2$-$(2^{n},2^{n-1}- 2^{n/2-1},(2^m-1)\cdot(2^{n-2} - 2^{n/2-1}))$ design. 
\end{definition}

\begin{remark}
	The designs $\D(F)$ for vectorial $(n,m)$-bent functions $F$ were introduced recently by Ding, Munemasa and Tonchev in~\cite{Ding19BCD}. Throughout the paper we will call these objects \emph{``addition design''}, motivated by terminology introduced by Bending in his thesis~\cite{Bending1993} for the designs $\D(f)$ of Boolean bent functions $f$ on $\F_2^n$. The term ``addition'' means, that blocks of the design $\D(f)$ are formed by supports of bent functions, obtained via addition of the original bent function $f\colon\F_2^n\rightarrow\F_2$ to those affine functions $l\colon\F_2^n\rightarrow\F_2$, which satisfy $\wt(f\oplus l)=2^{n-1}- 2^{n/2-1}$.
\end{remark}

\begin{remark}
	An incidence matrix of the addition design $\D(F)$ of an $(n,m)$-vectorial bent function, similarly to the Boolean case, can be constructed without the use of the linear code $\mathcal{C}(F)$. Recall that the \emph{dual} of a Boolean bent function $f\colon \F_2^n\rightarrow\F_2$ is a bent function $\tilde{f}\colon\F_2^n\rightarrow\F_2$, defined by $W_f(\aBold)=2^{n/2}(-1)^{\tilde{f}(\aBold)}$. Bending in~\cite[Theorem 9.6]{Bending1993} proved, that an incidence matrix of the design $\D(f)$ can be constructed with the help of the dual function $\tilde{f}$ in the following way (without loss of generality we assume, that $f(\mathbf{0})=0$):
	\begin{equation}
	\begin{split}
	M(\D(f))=& \ (m_{\xBold,\yBold})_{\xBold,\yBold\in\F_2^n},\mbox{ where}\\ m_{\xBold,\yBold}=& \ \tilde{f}(\xBold)\oplus f(\yBold)\oplus \langle\xBold,\yBold\rangle_n\oplus\tilde{f}(\mathbf{0}).
	\end{split}
	\end{equation}
	In this way, an incidence matrix of the addition design $\D(F)$ of an $(n,m)$-bent function $F$ (w.l.o.g. we assume $F(\mathbf{0})=\mathbf{0}$) can be constructed as the concatenation of incidence matrices of addition designs $\D(F_{\mathbf{b}})$ of the nonzero component functions $F_{\mathbf{b}}$ of $F$, namely:
	\begin{equation}
	M(\D(F))=\begin{bmatrix}
	M(\D(F_{\mathbf{b}_1}))\\
	M(\D(F_{\mathbf{b}_2}))\\
	\vdots\\
	M(\D(F_{\mathbf{b}_{2^n-1}}))
	\end{bmatrix}.
	\end{equation}
\end{remark}

\noindent Recently, Ding, Munemasa and Tonchev conjectured~\cite[Note 24]{Ding19BCD}, that extended-affine equivalence of vectorial bent functions, similarly to the Boolean case~\cite{Bending1993,DillonSchatz1987}, coincides with the isomorphism of their addition designs. In the following theorem we show that this conjecture is true.

\begin{theorem}
	Let $F$ and $F'$ be two $(n,m)$-bent functions. Bent functions $F$ and $F'$ are extended-affine equivalent if and only if addition designs $\mathbb{D}(F)$ and $\mathbb{D}(F')$ are isomorphic. 
	\begin{proof}
		First, let us recall the definition of the \emph{CCZ-equivalence} (abbreviation from Carlet-Charpin-Zinoviev). Two $(n,m)$-functions $F$ and $F'$ are called \emph{CCZ-equivalent}, if their graphs $\graph{F}$ and $\graph{F'}$ are affine equivalent, i.e. there exists an affine permutation $A$ of $\F_{2}^{n} \times \F_{2}^{m}$ s.t. $A\left(\graph{F}\right)=\graph{F'}$. The CCZ-equivalence is known as the most general equivalence relation for $(n,m)$-functions, however as it was shown in~\cite{BudaghyanC2009,KyureghyanP2008} two $(n,m)$-bent functions $F,F'$ are extended-affine equivalent if and only they are CCZ-equivalent. By \cite[Theorem 6.2]{browning2009apn} functions $F$ and $F'$ are CCZ-equivalent if and only if the linear codes $\mathcal{C}(F)$ and $\mathcal{C}(F')$ are equivalent. The proof of the statement now follows from \cite[Corollary 14]{Ding19BCD}, since linear codes $\mathcal{C}(F)$ and $\mathcal{C}(F')$ of $(n,m)$-bent functions $F$ and $F'$ are equivalent if and only if the addition designs $\mathbb{D}(F)$ and $\mathbb{D}(F')$ are isomorphic, since an incidence matrix of the addition design $\D(F)$ is a generator matrix of the code $\mathcal{C}(F)$.
	\end{proof}
\end{theorem}

\section{Translation designs of Boolean bent functions}
\label{section: 3 Translation designs}In this section we prove that isomorphism of translation designs $\dev(G_f)$ and $\dev(G_{f'})$ of Boolean bent functions $f,f'\colon\F_2^n\rightarrow\F_2$ is a coarser equivalence relation for Boolean bent functions than extended-affine equivalence. First, we give a general definition of the translation design.
\begin{definition}
	For a subset $A$ of an additive group $(G,+)$ the \emph{development} $\dev(A)$ of $A$ is an incidence structure, whose points are the elements in $G$, and whose blocks are the translates $A + g := \{a+g:a\in A \}$. For a Boolean function $f$ on $\F_2^n$ there are two ways to construct a translation design, see~\cite[Section 3]{Pott16}:
	\begin{itemize}
		\item $\dev(\support{f})$, which is a $2$-$(2^{n},2^{n-1} \pm 2^{n/2-1},2^{n-2} \pm 2^{n/2-1})$ design for a bent function $f$  on $\F_2^n$, with the  ``$+$'' sign if $f(\mathbf{0})=1$, and ``$-$'' otherwise;
		\item $\dev(\graph{f})$, which is a $\left(2^{n}, 2, 2^{n}, 2^{n-1}\right)$ divisible design for a bent function $f$ on $\F_2^n$.
	\end{itemize}
	It seems, there is no proper generalization of the translation designs $\dev(\support{f})$ for vectorial bent functions, while the second design $\dev(\graph{f})$ is defined in the same way. Thus, the translation design of an $(n,m)$-function $F$ is defined as:
	\begin{itemize}
		\item $\dev(\graph{F})$, which is a $\left(2^{n}, 2^{m}, 2^{n}, 2^{n-m}\right)$ divisible design for an $(n,m)$-bent function $F$.
	\end{itemize}
\end{definition}

\begin{remark}
	Despite translation and addition designs $\dev(\support{f})$ and $\D(f)$ of a Boolean bent function $f$ on $\F_2^n$ have the same parameters (up to a complement), in general they are non-isomorphic. However, for a quadratic bent function $f\colon\F_2^n\rightarrow\F_2$ the designs $\dev(\support{f})$ and $\D(f)$ are isomorphic, see~\cite[Theorem 11.9]{Bending1993}.
\end{remark}	
Further we denote by $\mathbf{J}_{2^n}$ the \emph{all-one-matrix} of order $2^n$ and by $A \otimes B$ the \emph{Kronecker product} of matrices $A$ and $B$. In the following proposition we observe, that from isomorphism of designs $\dev(\support{f})$ and $\dev(\support{f'})$ of Boolean (not necessarily bent) functions $f,f'$ on $\F_2^n$ follows the isomorphism of designs $\dev(\graph{f})$ and $\dev(\graph{f'})$. 
\begin{proposition}\label{proposition: Big designs from small}
	Let $f,f'\colon\F_2^n\rightarrow\F_2$ be two Boolean functions. If $\dev(\support{f})$ and $\dev(\support{f'})$ are isomorphic, then $\dev(\graph{f})$ and $\dev(\graph{f'})$ are isomorphic too.
	\begin{proof}
		First, we denote the complement of a Boolean function $f$ by $\bar{f}:= f\oplus 1$ and by $M_f$ an incidence matrix of the translation design $\dev(\graph{f})$, which can be computed as follows $M_f:=(f(\xBold\oplus\yBold))_{\xBold,\yBold\in\F_2^n}$, see~\cite{Weng20071096}. With the use of incidence matrices $M_{f}$ and $M_{\negation{f}}$ of translation designs $\dev(\support{f})$ and $\dev(\support{\bar{f}})$, respectively, one can decompose the incidence matrix $M(\dev(\graph{f}))$ of a Boolean function $f\colon\F_2^n\rightarrow\F_2$ in the following way  \cite{PolujanPott19DCC}:
		
		\begin{equation*}
		M(\dev(\graph{f}))=
		\left(
		\begin{array}{cc}
		M_{f} & M_{\negation{f}} \\
		M_{\negation{f}} & M_{f} \\
		\end{array}
		\right).
		\end{equation*}
		
		\noindent Since $\dev(\support{f})$ and $\dev(\support{f'})$ are isomorphic, there exist permutation matrices $P$ and $Q$, such that $M_{f}=P\cdot M_{f'}\cdot Q$. Clearly, $\dev(\support{\negation{f}})$ and $\dev(\support{\negation{f'}})$ are isomorphic with the same permutation matrices $P$ and $Q$, as one can see from the following calculations
		\begin{equation*}
		\begin{split}
		M_{\negation{f}}= & \ M_{f}\oplus\JJ_{2^n}=P\cdot M_{f'} \cdot Q \oplus\JJ_{2^n}
		\\= & \ P\cdot(M_{f'}\oplus\JJ_{2^n})\cdot Q=P\cdot M_{\negation{f'}}\cdot Q.
		\end{split}
		\end{equation*}
		
		\noindent Finally, since $M(\dev(\graph{f}))=(\II_2\otimes P)\cdot M(\dev(\graph{f'})) \cdot (\II_2\otimes Q)$, we conclude that $\dev(\graph{f})$ and $\dev(\graph{f'})$ are isomorphic.
	\end{proof}
\end{proposition}

\begin{remark}
	The converse of the previous statement is not true in general. A simple argument to see it, is that the design $\dev(\support{f})$ of a Boolean function $f$ on $\F_2^n$ is invariant for affine equivalence~\cite{Weng20071096}, that is $f(\xBold)=f'(\xBold A\oplus \bBold)$ for a non-degenerate $n\times n$ matrix $A$, but not extended-affine equivalence~\cite[Example 9.3.28]{KholoshaPott2013}. In general, there are many examples of non-isomorphic translation designs $\dev(\support{f})$ and $\dev(\support{f\oplus l})$, obtained by addition of an affine (and even linear) function $l$ to a bent function $f$ on $\F_2^n$, as it was mentioned by Dempwolff to the second author of this paper in a private communication. At the same time, the design $\dev(\graph{F})$ of an $(n,m)$-function $F$ is invariant for CCZ-equivalence and, hence, extended-affine equivalence, see~\cite{EdelP09}. In view of this remark we define isomorphic $(n,m)$-functions in the following way.
\end{remark}

\begin{definition}\label{definition: Isomorphic bent functions}
	Two $(n,m)$-functions $F,F'$ are \emph{isomorphic}, if translation designs $\dev(\graph{F})$ and $\dev(\graph{F'})$ are isomorphic.
\end{definition}

\begin{example}\label{example: Inequivalent but isomorphic bents}
	Let $f$ be a quadratic and $f'$ be a cubic Maiorana-McFarland bent functions on $\F_2^6$, given by their ANFs
	\begin{equation*}\label{equation: f and g}
	\begin{split}
	f(\xBold)=& \ x_1 x_2 \oplus x_3 x_4 \oplus x_5 x_6,\\
	f'(\xBold)=& \ x_1 x_2 \oplus x_3 x_4 \oplus x_5 x_6 \oplus x_1 x_3 x_5.
	\end{split}
	\end{equation*}
	
	\noindent Edel and Pott in~\cite[Example 1]{DBLP:conf/ima/EdelP09} observed that the designs $\dev(\support{f})$ and $\dev(\support{f'})$ are isomorphic. By Proposition~\ref{proposition: Big designs from small} the divisible designs $\dev(\graph{f})$ and $\dev(\graph{f'})$ are isomorphic too, and hence the functions $f$ and $f'$ are isomorphic in the sense of Definition~\ref{definition: Isomorphic bent functions}.
\end{example}
In the following proposition we show that using the direct sum construction one can always extend a pair of isomorphic incidence structures derived from Boolean and vectorial functions to an infinite family.
\begin{proposition}\label{proposition: Isomorphisms of functions and their extensions}
	Let $f,f'\colon\F_2^n\rightarrow\F_2$ be two Boolean functions and let $F,F'$ be two $(n,m)$-functions.
	\begin{enumerate}
		\item Let $h$ be a Boolean function on $\F_2^k$. If translation designs $\dev(\support{f})$ and $\dev(\support{f'})$ are isomorphic, then the translation designs $\dev(\support{f\oplus h})$ and $\dev(\support{f'\oplus h})$ of functions $f\oplus h$ and $f'\oplus h$ on $\F_2^{n}\times\F_2^{k}$ are isomorphic too.
		\item Let $H$ be a $(k,m)$-function. If translation designs $\dev(\graph{F})$ and $\dev(\graph{F'})$ are isomorphic, then the translation designs $\dev(\graph{F\oplus H})$ and $\dev(\graph{F'\oplus H})$ of $(n+k,m)$-functions $F\oplus H$ and $F'\oplus H$ are isomorphic too.
	\end{enumerate}
	\begin{proof}
		\emph{1.} Let $\xBold,\yBold\in\F_2^n$ and $\wBold,\zBold\in \F_2^k$. For any fixed $\wBold,\zBold\in \F_2^k$ the entry of the incidence matrix $M_{f\oplus h}$ of the translation design $\dev(\support{f\oplus h})$ labeled by $((\xBold,\wBold), (\yBold,\zBold))$ is $f(\xBold \oplus \yBold) \oplus h(\wBold \oplus \zBold)$. In this way, the incidence matrix $M_{f\oplus h}$ has the following form
		
		\begin{equation*}
		M_{f\oplus h}=(\JJ_{2^k}\otimes M_f) \oplus (M_h\otimes\JJ_{2^n}).
		\end{equation*}
		Since $\dev(\support{f})$ and $\dev(\support{f'})$ are isomorphic, there exist permutation matrices $P$ and $Q$, such that $M_{f}=P\cdot M_{f'}\cdot Q$. Finally, from the following equality
		
		\begin{equation*}
		M_{f\oplus h}=\left(\II_{2^k}\otimes P\right) \cdot M_{f'\oplus h} \cdot \left(\II_{2^k}\otimes Q\right)
		\end{equation*} 
		one can see that that designs $\dev(\support{f\oplus h})$ and $\dev(\support{f'\oplus h})$ are isomorphic.
		
		\noindent \emph{2.} Let $\aBold,\bBold\in\F_2^n$, $\cBold,\dBold\in\F_2^k$ and $\eBold,\fBold\in\F_2^m$. The point $(\aBold,\cBold,\eBold)$ is incident to the block $(\bBold,\dBold,\fBold)\oplus\graph{F\oplus H}$ of the translation design $\dev(\graph{F\oplus H})$ if and only if the point $(\aBold,\eBold)$ is incident to the block $(\bBold,\fBold \oplus H(\cBold\oplus\dBold))\oplus\graph{F}$ of the translation design $\dev(\graph{F})$. The statement now follows from the fact that for a block $B$ the mapping $\rho:B \mapsto B \oplus (\mathbf{0},H(\cBold\oplus\dBold))$ is an automorphism of designs $\dev(\graph{F})$ and $\dev(\graph{F'})$, which are isomorphic.
	\end{proof}
\end{proposition}
Further we show, that isomorphism of divisible designs for Boolean bent functions is a coarser equivalence relation than extended-affine equivalence.
\begin{theorem}\label{theorem: Inequivalent but isomorphic bent functions}
	Boolean bent functions, which are extended-affine inequivalent but isomorphic exist on $\F_2^n$ for all $n\ge 6$.
	\begin{proof}
		Let $g$ be a quadratic bent function on $\F_2^k$ and let $f$ and $f'$ be bent functions from the Example~\ref{example: Inequivalent but isomorphic bents}. By Proposition~\ref{proposition: Isomorphisms of functions and their extensions} Boolean functions $f\oplus g$ and $f'\oplus g$ on $\F_2^n$ with $n=k+6$ are isomorphic. Clearly, direct sums $f\oplus g$ and $f'\oplus g$ are bent, since all the functions $f,f'$ and $g$ are bent. Finally, since $\deg(f\oplus g)=2$ and $\deg(f'\oplus g)=3$, we get that functions $f\oplus g$ and $f'\oplus g$ are extended-affine inequivalent on $\F_2^n$.
	\end{proof}
\end{theorem}
\begin{remark}
	Extended-affine inequivalent Boolean bent functions $f$ and $f'$ on $\F_2^6$ from Example~\ref{example: Inequivalent but isomorphic bents} define isomorphic designs $\dev(\support{f})$ and $\dev(\support{f'})$ with 2-transitive automorphism group. According to Kantor~\cite[Theorem 1]{Kantor1985} any $2$-$(2^{n},2^{n-1}- 2^{n/2-1},2^{n-2} - 2^{n/2-1})$ design with a 2-transitive automorphism group is unique up to isomorphism. In general, if a design has a large automorphism group, it is more likely that it can be represented by several inequivalent difference sets (bent functions) due to the large symmetry. In this way, one may think that the reason why functions from Example~\ref{example: Inequivalent but isomorphic bents} have isomorphic translation designs is the 2-transitivity of the automorphism group. In the following example we show that isomorphic translation designs $\dev(\support{f})$ and $\dev(\support{f'})$ of EA-inequivalent bent functions $f$ and $f'$ do not necessarily need to have a 2-transitive automorphism group.
\end{remark}

\begin{example}\label{example: Inequivalent but isomorphic bents 2}
	Let $f,f'$ be two Maiorana-McFarland bent functions on $\F_2^{10}$ given by	
	\begin{equation*}
	\begin{split}
	f(\xBold) = & \ x_1 x_6 \oplus x_2 x_7 \oplus x_3 x_8\oplus x_4 x_9 \oplus x_5 x_{10} \\
	\oplus& \  x_1 x_2 x_3 x_4 x_5, \\
	f'(\xBold) = & \ f(\xBold) \oplus x_4 \oplus x_6 \oplus x_8 \oplus x_{10} \oplus x_1 x_2 \oplus x_2 x_3 \\
	\oplus & \ x_1 x_2 x_3 \oplus x_2 x_4 x_5 \oplus x_1 x_2 x_4 x_5 \oplus x_2 x_3 x_4 x_5.
	\end{split}
	\end{equation*}
	With Magma~\cite{MR1484478} one can check that $|\Aut(\mathcal{C}(f))|=2^{30}\cdot 3^2\cdot5\cdot 7\cdot 31$ and $|\Aut(\mathcal{C}(f'))|=2^{30}\cdot 3^2\cdot 7$, what implies that functions $f$ and $f'$ are extended-affine inequivalent. However, the designs $\dev(\support{f})$ and $\dev(\support{f'})$ are isomorphic. First, we observe that in general designs $\dev(\support{f})$ and $\dev(\support{f'})$ are isomorphic if and only if there exist a pair of permutations $\pi,\sigma\colon\F_2^n\rightarrow\F_2^n$, such that $f(\pi(\xBold)\oplus\sigma(\yBold))=f'(\xBold\oplus\yBold)$ holds for all $\xBold,\yBold \in \F_2^n$, since an incidence matrix $M_f$ of the translation design $\dev(\support{f})$ can be computed as $M_f:=(f(\xBold\oplus\yBold))_{\xBold,\yBold\in\F_2^n}$. It is easy to check, that the following nonlinear functions $\pi,\sigma\colon\F_2^{10}\rightarrow\F_2^{10}$, given by  algebraic normal forms
	\begin{equation*}
	\begin{split}
	\pi(\xBold)=& \ (x_1, x_2, x_3, x_4, x_1  \oplus  x_5, x_1  \oplus  x_{10}  \oplus  x_2 x_3  \oplus  x_5  \oplus  x_6, \\
	& \ \phantom{(}x_1 x_3  \oplus  x_7, x_1 x_2  \oplus  x_8, x_9, x_1  \oplus  x_{10}),\\
	\sigma(\yBold) = & \ \yBold \oplus (1, 0, 1, 0, y_1, y_1 \oplus y_{10} \oplus y_2 \oplus y_2 y_3 \oplus y_5, \\
	&\ \phantom{(} y_1 \oplus y_3 \oplus y_1 y_3, y_2 \oplus y_1 y_2, 1 , 1 \oplus y_1),
	\end{split}
	\end{equation*}
	are permutations and satisfy $f(\pi(\xBold)\oplus\sigma(\yBold))=f'(\xBold\oplus\yBold)$ for all $\xBold,\yBold\in\F_2^{10}$. In this way, designs $\dev(\support{f})$ and $\dev(\support{f'})$ are isomorphic. Further we observe that the 2-rank of any $2$-$(2^{n},2^{n-1}- 2^{n/2-1},2^{n-2} - 2^{n/2-1})$ design $D$ (i.e. $\rank_{\F_2}M(D)$) with a 2-transitive automorphism group equals $n+2$. Any such a design is isomorphic to $\dev(\support{g})$ of a quadratic bent function $g$ on $\F_2^n$, and $\tworank (\dev(\support{g}))=n+2$ as it was shown in~\cite[Corollary 3.8]{Weng2008}. Since $f$ is a Maiorana-McFarland bent function of the form $\langle \xBold',\xBold'' \rangle_{n/2} \oplus h(\xBold'')$, where $\xBold',\xBold''\in\F_2^{n/2}$ and $h$ is a monomial function on $\F_2^{n/2}$ with $\deg(h)>3$, we have $\tworank(\dev(\support{f}))=n-2\deg(h) + 2^{\deg(h)}$ by \cite[Corollary 3.8]{Weng2008}. In this way, $\tworank(D)=12$, which is different from $ \tworank(\dev(\support{f}))=\tworank(\dev(\support{f'}))=32$, from what follows that automorphism groups of designs $\dev(\support{f})$ and $\dev(\support{f'})$ are not 2-transitive.
\end{example}
\section{Translation designs of vectorial bent functions}
\label{section: 4 Translation designs of vectorial bent functions}
In the previous section we showed that extended-affine inequivalent Boolean bent functions can give isomorphic translation designs. Further we show that the same phenomenon does not occur for the vectorial bent functions in 6 variables. We classify and enumerate all $(6,m)$-vectorial bent functions and show, that two vectorial bent functions $F$ and $F'$ in six variables are EA-equivalent if and only if their translation designs $\dev(\graph{F})$ and $\dev(\graph{F'})$ are isomorphic. 	
\subsection{Extension invariants of bent functions}
\label{subsection: 4.1 Extension invariants}
We denote by $\BBnm$ the \emph{set of all $(n,m)$-bent functions}, by $\Affine$ the \emph{set of all $(n,m)$-affine functions} and by $\ABnm$ the \emph{set of affine-free $(n,m)$-bent functions}, i.e. any $f\in\ABnm$ contains no affine terms in its ANF. Since bentness is invariant with respect to the addition of affine terms, the cardinalities of these three sets are related as follows $|\BBnm|=|\ABnm|\cdot |\Affine|$.

For the sake of convenience we denote by $C^{m}_i$ an $i$-th EA-equivalence class of $(n,m)$-bent functions.	On the set $\bigcup_{m=1}^{n/2}\BBnm$ we introduce the order relation ``$\prec$'' in the following way. Let $m<l$ and $C^{m}_i$ and $C^{l}_j$ be two equivalence classes of $(n,m)$- and $(n,l)$-bent functions, respectively.  We say that a function $F\in C^{m}_i$ \emph{is contained} in $G\in C^{l}_j$ and write $F\prec G$, if the first $m$ coordinate functions of $G(\xBold)=(g_1(\xBold),\ldots,g_l(\xBold))^T$ form a function $F$, that is $F(\xBold)=(g_1(\xBold),\ldots,g_m(\xBold))^T$. Similarly, we say that $F\in C^{m}_i$ \emph{is contained} in the equivalence class $C^{l}_j$ and write $F \prec C^{l}_j$, if there exist a representative $G\in C^{l}_j$, such that $F\prec G$. Finally, we say that the equivalence class $C^{m}_{i}$ \emph{is contained} in $C^{l}_{j}$ and denote it by $C^{m}_{i} \prec C^{l}_{j}$ if there exist $F\in C^{m}_{i}$, such that $F \prec C^{l}_{j}$.
\begin{definition}
	An $(n,m)$-bent function $F$ is called \emph{extendable}, if the there exists a Boolean bent function $f\colon\F_2^n\rightarrow\F_2$, such that the function $G\colon \xBold\in\F_2^n\mapsto \left(F(\xBold),f(\xBold)\right)^T$ is $(n,m+1)$-bent. If no such a bent function $f$ exists, the function $F$ is called \emph{non-extendable}.
\end{definition}
\begin{remark}
	The problem of the existence of non-extendable bent functions $F\colon\F_p^n\rightarrow\F_p^m$ has mostly been studied for the case $p$ odd, see~\cite[Section 4]{OzbudakPott2014}. The particular case of this problem, namely $p=2$ and $m=1$, is closely related to the Tokareva's conjecture~\cite[Hypothesis 1]{Tokareva2011}, that any Boolean function on $\F_2^n$ of degree at most $n/2$ can be represented as the sum of two Boolean bent functions on $\F_2^n$. For instance, a single example of a non-extendable Boolean bent function would disprove the Tokareva's conjecture.
\end{remark}	
\begin{definition}\label{definition: Bent friend of a function}
	Let $F$ be an $(n,m)$-bent function. Further we define the following two sets
	\begin{equation}\label{equation: Affine-free bent friends}
	\begin{split}
	\mathcal{F}(F):=& \ \left\{ f\in\AB{n}{1}\colon  (F,f)^T \mbox{ is } (n,m+1)\mbox{-bent} \right\}, \\ \ext(F):= & \ \{ (F,f)^T \colon f \in \mathcal{F}(F) \},
	\end{split}
	\end{equation}
	namely, $\mathcal{F}(F)$ is the set of affine-free Boolean bent functions, which can extend an $(n,m)$-bent function $F$ to an $(n,m+1)$-bent function and $\ext(F)$ is the set of extensions of a function $F$. Clearly, different extensions may lead to different equivalence classes. In this way, we define
	\begin{equation}
	\begin{split}
	\mathcal{F}(F,C^{m+1}_{j}):= & \ \left\{ f\in\AB{n}{1}\colon  (F,f)^T \in C^{m+1}_j \mbox{ is } \right. \\ 
	& \left. \ \ (n,m+1)\mbox{-bent} \right \}
	\end{split}
	\end{equation}
	as the set of affine-free Boolean bent functions, which can extend an $(n,m)$-bent function $F$ to the equivalence class $C^{m+1}_{j}$. Similarly, we define the set of extensions of the function $F$, which belong to the equivalence class $C^{m+1}_{j}$, that is
	\begin{equation}
	\ext(F,C^{m+1}_{j}):=\{ (F,f)^T \colon f \in \mathcal{F}(F,C^{m+1}_{j}) \}.
	\end{equation}
	Clearly, the collection of sets $\ext(F,C^{m+1}_{j})$ forms a partition of $\ext(F)$, namely
	\begin{equation}\label{equation: Bent friends}
	\ext(F)=\bigsqcup_{j:F \prec C^{m+1}_{j}} \ext(F,C^{m+1}_{j}).
	\end{equation}
\end{definition}
\begin{remark}
	Non-extendable $(n,m)$-bent functions $F$ are also called \emph{lonely}, see~\cite{Meidl2019BFA}. In this way, it is essential to call the following sets:
	\begin{itemize}
		\item $\mathcal{F}(F)$~--- the \emph{set of bent friends} of a bent function $F$;
		\item $\mathcal{F}(F,C^{m+1}_j)$~--- the \emph{set of bent friends} of $F$, \emph{leading to the equivalence class} $C^{m+1}_j$.
	\end{itemize}
	Indeed, according to Definition~\ref{definition: Bent friend of a function}, a bent function $F$ is lonely, if it has no bent friends, that is $|\mathcal{F}(F)|=0$. We also call $(n,n/2)$-bent functions \emph{absolutely non-extendable (lonely)}, since $(n,m)$-bent functions do not exist for $m>n/2$ due to the Nyberg bound~\cite{Nyberg91}.
\end{remark}	
\begin{definition}
	For an $(n,m+1)$-function $G$ we denote by $\mathcal{S}(G)$ a set of equivalence classes of $(n,m)$-bent functions $F,F'$ of the form $F\colon\xBold\in\F_2^n\mapsto A_F \circ G(\xBold)$ and $F'\colon\xBold\in\F_2^n\mapsto A_{F'} \circ G(\xBold)$, where $A_{F},A_{F'}\colon \F_2^{m+1}\rightarrow \F_2^{m}$ are surjective linear mappings and the equivalence relation ``$\sim$'' is defined as follows: $F \sim F'$ if $\{F_{\bBold}\colon \bBold\in\F_2^m \}=\{F'_{\bBold}\colon \bBold\in\F_2^m \}$, i.e. functions $F$ and $F'$ have the same component functions. We will call the set $\mathcal{S}(G)$ the \emph{set of different} $(n,m)$\emph{-bent spaces} of a function $G$. Clearly, its cardinality is given by $|\mathcal{S}(G)|={m+1 \atopwithdelims [ ] m}_{2}=2^{m+1}-1$. Finally, for an $(n,m)$-bent function $F\prec G$ we denote by $\mathcal{S}(F,G):=\{ H\in \mathcal{S}(G)\colon H \mbox{ is EA-equivalent to } F \}$ the set of different $(n,m)$-bent spaces of $G$, which are EA-equivalent to $F$.
\end{definition}

Further, we show that cardinalities of the sets $\mathcal{F}(F,C^{m+1}_{j})$ and $\mathcal{S}(F,G)$ do not depend on representatives of equivalence classes and thus are invariants for extended-affine equivalence.

\begin{proposition}\label{lemma: Bent friends are invariants} Let $F,F'\in C^{m}_{i} \prec C^{m+1}_{j}$ be two $(n,m)$-bent functions and $G,G'\in C^{m+1}_{j}$ be two $(n,m+1)$-bent functions. Then the following hold.
	\begin{enumerate}
		\item  $|\mathcal{F}(F,C^{m+1}_j)|=|\mathcal{F}(F',C^{m+1}_j)|$;
		\item If $F\prec G$ and $F'\prec G'$, then  $|\mathcal{S}(F, G)|=|\mathcal{S}(F', G')|$.
	\end{enumerate}
	\begin{proof}
		\emph{1.} Let $F$ and $F'$ be EA-equivalent, i.e. $F=A_1\circ F' \circ A_2 \oplus A_3$. Clearly, if $f$ is a bent friend of $F$, then $f':=f\circ A_2$ is a bent friend of $F'$. Moreover, the non-degenerate affine transformation $A_2$ maps different bent friends to different ones. 
		
		\noindent\emph{2.} Assume that $H\in\mathcal{S}(F,G)$, i.e. there exist non-degenerate linear mapping $A_H\colon \F_2^{m+1}\rightarrow \F_2^{m}$ such that $H=A_H \circ G=B_1\circ F\circ B_2 \oplus B_3$,
		since $H,F \in C^{m}_i$. Further, we may assume that $G'=A_1\circ G\circ A_2 \oplus A_3$, since $G,G'\in C^{m+1}_j$. Multiplying the latter equality by $A_H\circ A_1^{-1}$ from left and substituting it the second last, one gets $A_H\circ A_1^{-1} \circ G' = H\circ A_2\oplus A_H \circ A_1^{-1}\circ A_3$. Finally,
		denoting by $A_{H'}:=A_H\circ A^{-1}_1$, we get that the function $H':=A_{H'}\circ G'$ is EA-equivalent to $F'$, and hence $H'\in \mathcal{S}(F',G')$.
	\end{proof}
\end{proposition}
\noindent In this way, for two equivalence classes $C^{m}_i \prec C^{m+1}_j$ we denote by $|\mathcal{F}(C^{m}_i, C^{m+1}_j)|$, the number of Boolean bent functions, which can extend any representative of $C^{m}_i$ to the class $C^{m+1}_j$ and by $|\mathcal{S}(C^{m}_i, C^{m+1}_j)|$, the number of different bent spaces contained in $C^{m+1}_j$, which represent the equivalence class $C^{m}_i$, that is
\begin{equation}
\begin{split}
|\mathcal{F}(C^{m}_i,C^{m+1}_j)|:=& \ |\mathcal{F}(F,C^{m+1}_j)|\mbox{ and}\\ |\mathcal{S}(C^{m}_i,C^{m+1}_j)|:=& \ |\mathcal{S}(F,C^{m+1}_j)| \mbox{ for } F\in C^{m}_i.
\end{split}
\end{equation}
In the next subsection we will use the number of bent friends $	|\mathcal{F}(C^{m}_i,C^{m+1}_j)|$ in order to enumerate all vectorial bent functions in six variables and the number of bent spaces $|\mathcal{S}(C^{m}_i,C^{m+1}_j)|$ in order to verify these computations.
\subsection{Classification and enumeration of vectorial bent functions in six variables}
\label{subsection: 4.2 Classification and enumeration of vectorial bent functions in six variables}
Now we describe how to determine the cardinality of the equivalence class $C^{m+1}_{j}$, provided its structure is known.
\begin{proposition}\label{proposition: Cardinalities of equivalence classes}
	Let $C^{m}_1,\ldots, C^{m}_k \prec C^{m+1}_j$ be all equivalence classes of $(n,m)$-bent functions, contained in $C^{m+1}_j$. Then the cardinality of the class $C^{m+1}_j$ is equal to
	\begin{equation}\label{equation: Cardinality of bent class}
	|C^{m+1}_j|=2^{n+1}\cdot\sum_{i=1}^{k} |C^{m}_i|\cdot |\mathcal{F}(C^{m}_i,C^{m+1}_j)|.
	\end{equation}
	\begin{proof}
		Any function $G\in C^{m+1}_j$ can be considered as an extension of a function $F \in C^m_i \prec C^{m+1}_j$, that is  $G=(F,f)^T\in C^{m+1}_j$ for $f\in \mathcal{F}(F,C^{m+1}_{j})$. There are $k$ ways to select an equivalence class $C^{m}_i \prec C^{m+1}_j$, such that $F\in C^m_i$, and there are $|C^{m}_i|$ ways to choose a representative $F$. Finally, for any representative $F\in C^{m}_i$ there exist exactly $2^{n+1}\cdot|\mathcal{F}(C^{m}_i,C^{m+1}_j)|$ ways to extend it to a function $G \in C^{m+1}_j$, since bentness is invariant with respect to addition of affine terms.
	\end{proof}
\end{proposition}
Further we summarize the above ideas in the form of a recursive algorithm.
\begin{algorithm}
	\caption{Classification and enumeration of all $(n,m)$-bent functions}
	\label{algorithm: Classification and enumeration of all vectorial bent functions.}
	\begin{algorithmic}[1]
		\Require All pairs $(F^1_i\in C^1_i,|C^1_i|)$, where \Indent $\BB{n}{1}=\bigsqcup_{i} \{ f \colon f \in C^{1}_{i} \}$. \EndIndent
		\Ensure All pairs $(F^m_i\in C^m_i,|C^m_i|)$, where \Indent $\BB{n}{m}=\bigsqcup_{i} \{ f \colon f \in C^{m}_{i} \}$ for all $2\le m\le n/2$. \EndIndent
		\For{$m=1$ to $n/2-1$}
		\For{all equivalence classes $C^m_i$}
		\State \textbf{Construct} the set of extensions $\ext(F^m_i)$.
		\State \textbf{Classify} all $(n,m+1)$-bent functions from the set \Indent $\ext(F^m_i)$ by constructing  the partition \EndIndent $$\ext(F^m_i)=\bigsqcup_{j_i:F^m_i \prec C^{m+1}_{j_i}}\ext(F^m_i,C^{m+1}_{j_i}).$$
		\State \textbf{Compute} the number of bent friends $$|\mathcal{F}(C^{m}_i,C^{m+1}_{j_i})|:=|\mathcal{F}(F^m_i,C^{m+1}_{j_i})|.$$
		\EndFor
		\State \textbf{Identify} all equivalent classes $C^{m+1}_{j_i}$ with the class \Indent $C^{m+1}_{j}$ and set $F^{m+1}_{j}$ to be a random representative \EndIndent of the equivalence class $C^{m+1}_{j}$.
		\State \textbf{Compute} the numbers of bent spaces $$|\mathcal{S}(C^{m}_i,C^{m+1}_j)|:=|\mathcal{S}(C^{m}_i,C^{m+1}_{j_i})|$$ \Indent and cardinalities of equivalence  classes \EndIndent $$|C^{m+1}_j|=2^{n+1} \cdot\sum_{i=1}^{k}  |C^{m}_i|\cdot |\mathcal{F}(C^{m}_i,C^{m+1}_{j_i})|.$$ 
		\EndFor
		\State \textbf{Return} pairs $(F^m_i\in C^m_i,|C^m_i|)$ for all $2\le m\le n/2$.
	\end{algorithmic}
\end{algorithm}	
Applying the Algorithm~\ref{algorithm: Classification and enumeration of all vectorial bent functions.} for Boolean bent functions in six variables, we obtain the main result of this section.
\begin{theorem}
	For vectorial bent functions in $6$ variables the following hold.
	\begin{enumerate}
		\item There are $23{,}392{,}233{,}361{,}244{,}160\approx2^{54{.}37}$ vectorial $(6,2)$-bent functions, which are divided into $9$ extended-affine equivalence classes.
		\item There are $121{,}282{,}113{,}886{,}947{,}901{,}440\approx2^{66{.}71}$ vectorial $(6,3)$-bent functions, which are divided into $13$ extended-affine equivalence classes.
	\end{enumerate}
	Moreover, if a $(6,m)$-bent function $F$ is non-extendable, then $F$ is absolutely non-extendable, i.e. it has $m=3$.
	\begin{proof}
		Further we discuss the main steps of the Algorithm~\ref{algorithm: Classification and enumeration of all vectorial bent functions.} and explain how one can verify our computational results.
		
		\noindent\emph{Input.} For the input of the Algorithm~\ref{algorithm: Classification and enumeration of all vectorial bent functions.} one has to provide the pairs $(F^1_i \in C^1_i,|C^1_i|)$ for all equivalence classes $C^1_i$, which form the partition of the set of Boolean bent functions $\BB{6}{1}$. The representatives of 4 equivalence classes are well-known and could be found in~\cite{ROTHAUS1976300}. For the cardinalities of the equivalence classes we refer to~\cite[Table 8.7]{Preneel1993}.
		
		\noindent\emph{Output.} For the computation of the collections $\mathcal{F}(F^m_i)$ one first has to construct all affine-free Boolean bent functions $\AB{6}{1}$, which can be efficiently listed as described in~\cite{LangevinL11,Meng20085576}. Further, for a given representative $F^m_i\in C^m_i$ we construct the set $\mathcal{F}(F^m_i)$, by checking directly the characteristic property in~\eqref{equation: Affine-free bent friends}. The classification of functions $G\in \ext(F^m_i)$ is carried out with Magma~\cite{MR1484478}, by checking equivalence of linear codes $\mathcal{C}(G)$ introduced in Definition~\ref{definition: Addition designs}.
		
		In this way, Algorithm~\ref{algorithm: Classification and enumeration of all vectorial bent functions.} constructs $n/2-1$ layers of the weighted Hasse diagram, given in Figure~\ref{figure: Hasse Diagram} as follows. For all $2\le m\le n/2-1$ we draw an edge between equivalence classes $C^{m}_i$ and $C^{m+1}_j$ if $C^{m}_i \prec C^{m+1}_j$ and assign two weights with it. The first number closer to the equivalence class $C^{m}_i$ is the number of bent spaces $|\mathcal{S}(C^{m}_i,C^{m+1}_j)|$ and the second number, closer to $C^{m+1}_j$, is the number of bent friends $|\mathcal{F}(C^{m}_i,C^{m+1}_j)|$. Note that, if $C^{m}_1,\ldots, C^{m}_k \prec C^{m+1}_j$ are all equivalence classes, contained in $C^{m+1}_j$, then the following relation holds 
		\begin{equation*}
		\sum_{i=1}^{k}|\mathcal{S}(C^{m}_i,C^{m+1}_j)|= {m+1 \atopwithdelims [ ] m}_{2}=2^{m+1}-1.
		\end{equation*}
		In Figure~\ref{figure: Hasse Diagram} we list exact cardinalities $|C^1_i|$ for all equivalence classes $C^1_i$,  while for equivalence classes $C^{m\ge 2}_i$, due to the lack of a space, we give only approximate values. Note that, the exact values $|C^{m\ge 2}_i|$ can be recovered with the Proposition~\ref{proposition: Cardinalities of equivalence classes}.
		\begin{figure*}[tp]
			\centering
			\caption{The structure of equivalence classes $C^{m}_i$ of vectorial bent functions in $6$ variables}
			\label{figure: Hasse Diagram}
			\begin{center}	\includegraphics[width=\textwidth]{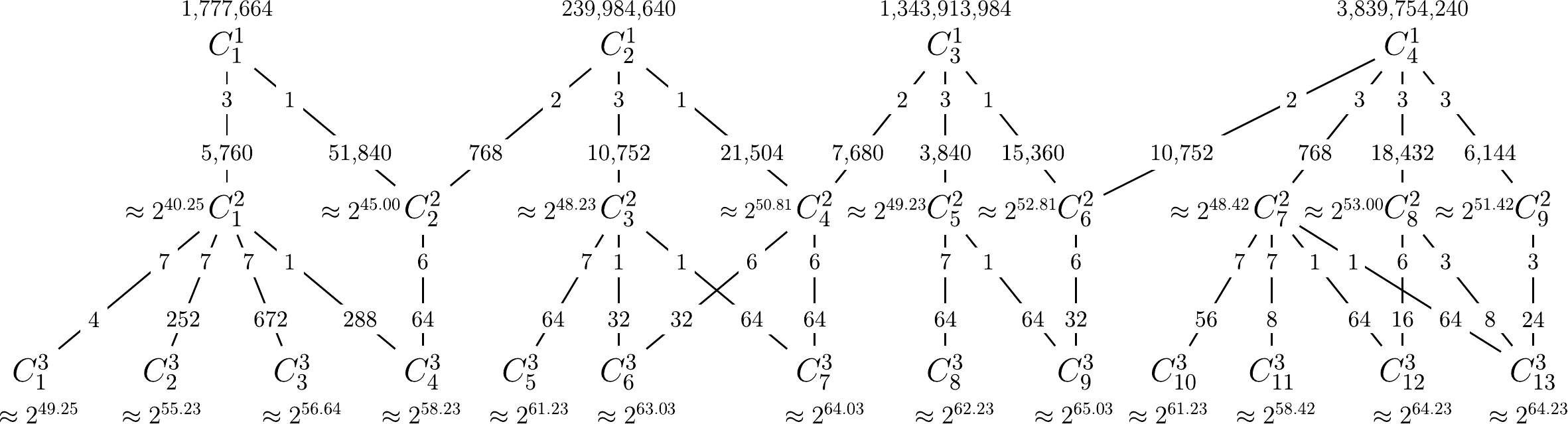}
			\end{center}
		\end{figure*}
		
		\noindent Finally, we give the total number of bent functions in six variables in Table~\ref{table: Count of bent functions in 6 variables} and provide algebraic normal forms of representatives of the equivalence classes together with their invariants in Appendix~\ref{appendix: A}.
		\begin{table*}[tp]
			\centering
			\caption{Classification and count of bent functions in 6 variables}
			\label{table: Count of bent functions in 6 variables}
			\scalebox{1}{
				\begin{tabular}{|c|r|r|c|c|}
					\hline
					$(n,m)$ & \multicolumn{1}{c|}{$|\ABnm|$}     & \multicolumn{1}{c|}{$|\BBnm|=|\ABnm|\cdot2^{m(n+1)}$} & \multicolumn{1}{c|}{$\#$ Eq. cl.} \\ \hline
					$(6,1)$ & $48{,}386{,}176\approx2^{25{.}33}$ & $5{,}425{,}430{,}528\approx2^{32{.}33}$ & 4 \\
					$(6,2)$ & $1{,}427{,}748{,}618{,}240\approx2^{40{.}37}$ &$23{,}392{,}233{,}361{,}244{,}160\approx2^{54{.}37}$ & 9 \\
					$(6,3)$ & $57{,}831{,}818{,}526{,}720\approx2^{45{.}71}$ & $121{,}282{,}113{,}886{,}947{,}901{,}440\approx2^{66{.}71}$ & 13 \\
					\hline
			\end{tabular}}
		\end{table*}	
		
		\noindent\emph{Verification.} First, one may observe that cardinalities of all equivalence classes and hence of the sets $\BB{6}{m}$ are divisible by the order of the general linear group $\mbox{GL}(m,2)$, which is given by $|\mbox{GL}(m,2)|=\prod _{k=0}^{m-1} \left(2^m-2^k\right)$. We also observe that the number of affine-free quadratic $(6,2)$-bent functions established with Algorithm~\ref{algorithm: Classification and enumeration of all vectorial bent functions.} coincides with the theoretically computed value $|\AB{n}{2}|$, given in~\cite[Theorem 1]{PottSZ16}. Further we note that for equivalence classes $C^{m}_{i},C^{m}_{i'} \prec C^{m+1}_j$, contained in $C^{m+1}_j$ the following relation holds
		\begin{equation*}
		\frac{\left|C^{m}_{i}\right| \cdot|\mathcal{F}(C^{m}_{i},C^{m+1}_{j})|}{|\mathcal{S}(C^{m}_{i},C^{m+1}_{j})|}=\frac{\left|C^{m}_{i'}\right| \cdot|\mathcal{F}(C^{m}_{i'},C^{m+1}_{j})|}{|\mathcal{S}(C^{m}_{i'},C^{m+1}_{j})|},
		\end{equation*}
		\noindent since the portion of $(n,m)$-bent functions from the class $C^m_i$, contained in the equivalence class $C^{m+1}_j$ equals to
		$$\frac{|\mathcal{S}(C^{m}_{i},C^{m+1}_{j})|}{(2^{m+1}-1)\cdot 2^{n+1} \cdot |\mathcal{F}(C^{m}_{i},C^{m+1}_{j})|}.$$
		\noindent Finally, from Figure~\ref{figure: Hasse Diagram} one can see that the only non-extendable bent functions in 6 variables are those, which achieve the Nyberg bound, i.e. $(6,3)$-bent functions.
	\end{proof}
\end{theorem}

We also checked that the only equivalence classes of Boolean bent functions, which lead to isomorphic translation designs are $C^1_1$ and $C^1_2$, as one can see from Example~\ref{example: Inequivalent but isomorphic bents} and Table~\ref{table: Invariants of (6,1)-bent functions}. Surprisingly, in contrast to the Boolean case, one cannot construct isomorphic translation designs from extended-affine inequivalent vectorial bent functions.
\begin{theorem}\label{theorem: Isomorphic functions are equivalent}
	Let $F$ and $F'$ be two $(6,m)$-bent functions with $m\ge2$. The following statements are equivalent.
	\begin{enumerate}
		\item Bent functions $F$ and $F'$ are extended-affine equivalent.
		\item Divisible designs $\dev(\graph{F})$ and $\dev(\graph{F'})$ are isomorphic.
	\end{enumerate}
	\begin{proof}
		All computations about equivalence and isomorphism are carried out with Magma~\cite{MR1484478}. Invariants of equivalence classes and their translation designs are listed in Table~\ref{table: Invariants of (6,2)-bent functions} and Table~\ref{table: Invariants of (6,3)-bent functions}.
	\end{proof}
\end{theorem}

\begin{remark}
	It is well-known, that all Boolean bent functions in six variables up to EA-equivalence can be described by two classical constructions: Maiorana-McFarland $\mathcal{M}$, and Desarguesian partial spread $\mathcal{PS}_{ap}$, which have straightforward generalizations to the vectorial case, see~\cite[p. 309]{Mesnager2016}. We endow $\F_{2}^{n/2}$ with the structure of the finite field $\left(\F_{2^{n/2}},+,\cdot \right)$ and identify $\F_{2}^{n}$ with $\F_{2^{n/2}} \times \F_{2^{n/2}}$. The \emph{strict Maiorana-McFarland class} $\mathcal{M}$ of vectorial bent functions is the set of $(n,m)$-functions $F$ of the form $F(x,y)=L(x\cdot \pi (y))+G(y)$, where $L\colon \F_{2^{n/2}} \rightarrow \F_{2^{m}}$ is a linear or an affine function, $\pi \colon \F_{2^{n/2}} \rightarrow \F_{2^{n/2}}$ is a permutation, and $G\colon \F_{2^{n/2}} \rightarrow \F_{2^{m}}$ is an arbitrary $(n/2,m)$-function. The $\mathcal{PS}_{ap}$ \emph{class} of vectorial bent functions is the set of $(n,m)$-bent functions $F$ of the form $F(x, y):=H\left(x\cdot y^{2^{n/2}-2}\right)=H\left(x/y\right)$ with $x/y=0$ if $ y=0$ for $x, y \in \F_{2^{n/2}}$ and $H$ is
	a balanced $(n/2, m)$-function (or, equivalently, permutation if $m=n/2$).
	
	In the following table we list equivalence classes $C^m_i$ of $(6,3)$-bent functions, which can be described by $\mathcal{M}$ and $\mathcal{PS}_{ap}$ classes. Note that $(6,2)$-bent functions from $\mathcal{M}$ and $\mathcal{PS}_{ap}$ can be constructed as proper bent subspaces of $(6,3)$-bent from $\mathcal{M}$ and $\mathcal{PS}_{ap}$ classes.

	\begin{table}[H]
		\caption{Equivalence classes of $(6,3)$-bent functions, described by classical constructions}
		\label{table: Classical constructions of vectorial bent functions}
		\begin{subtable}{1\linewidth}
			\caption{Balanced mappings on $\F_2^3$}
			\label{table: Balanced vectorial functions}
			\centering
			\scalebox{1}{
				\begin{tabular}{|c|c|}
					\hline
					$\pi_{i}$ & $z\mapsto \pi_i(z)$                         \\ \hline
					$\pi_{1}$ & $z$                         \\
					$\pi_{2}$ & $z^3$                     \\
					$\pi_{3}$ & $z + z^3 + z^5$             \\
					$\pi_{4}$ & $z^2 + z^3 + z^4 + z^5 + z^6$ \\ \hline
				\end{tabular}
			}
		\vspace{0.25cm}
		\end{subtable}
		
		\begin{subtable}{1\linewidth}
			\caption{$\mathcal{M}$ class}
			\label{table: MM vectorial bent functions}
			\centering
			\scalebox{1}{
				\begin{tabular}{|c|c|}
					\hline
					$C^3_{i\phantom{0}}$   & $x \cdot \pi_i(y) + G(y)$ \\ \hline
					$C^3_{1\phantom{0}}$ & $x \cdot \pi_1(y) $ \\
					$C^3_{2\phantom{0}}$ & $x \cdot \pi_1(y) + (y + y^2 + y^3 + y^6)$ \ \\
					$C^3_{4\phantom{0}}$ & $x \cdot \pi_1(y) + (y^3 + y^5 + y^6 + y^7) $ \\ \hline
					$C^3_{5\phantom{0}}$ & $x \cdot \pi_4(y)$ \\ \hline
					$C^3_{8\phantom{0}}$ & $x \cdot \pi_3(y)$ \\ \hline
					$C^3_{10}$ & $x \cdot \pi_2(y) + (y + y^2 + y^3 + y^6) \ $ \\ 
					$C^3_{11}$ & $x \cdot \pi_2(y)$ \\ \hline
				\end{tabular}	
			}
		\vspace{0.25cm}
		\end{subtable}
		
		\begin{subtable}{1\linewidth}
			\caption{$\mathcal{PS}_{ap}$ class}
			\label{table: PSap vectorial bent functions}
			\centering
			\scalebox{1}{
				\begin{tabular}{|c|c|}
					\hline
					$C^3_{i}$  & $H(x/y)$ \\ \hline
					$C^3_{11}$ & $\pi_1(x/y)$ \\
					$C^3_{12}$ & $\pi_3(x/y)$ \\
					$C^3_{13}$ & $\pi_4(x/y)$ \\ \hline
				\end{tabular}
			}
		\end{subtable}
	\end{table}
	In this way, from Figure~\ref{figure: Hasse Diagram} and Table~\ref{table: Classical constructions of vectorial bent functions} one can see that the only ``missing'' equivalence classes of $(6,3)$-bent functions are $C^{3}_{3},C^{3}_{6},C^{3}_{7},C^{3}_{9}$ and of $(6,2)$-bent functions are $C^{2}_{4},C^{2}_{6}$. In view of this observation we conclude, that in contrast to the Boolean case, vectorial versions of the classical Maiorana-McFarland and Desarguesian partial spread constructions do not cover the whole set of vectorial bent functions in six variables. 
\end{remark}

\section{Summary and concluding remarks}
\label{section: 5 Conclusion and future work}
In this paper we compared different concepts of equivalence relations for Boolean and vectorial bent functions: extended-affine equivalence of functions, isomorphism of translation designs and isomorphism of addition designs. We summarize our results in the following table.
\begin{table}[H]
	\caption{EA-equivalence vs. isomorphism of designs for bent functions}
	\label{table: Equivalence vs isomorphism}
	\centering
	\scalebox{1}{\begin{tabular}{|>{\centering\arraybackslash}m{.34\linewidth}|>{\centering\arraybackslash}m{.26\linewidth}|>{\centering\arraybackslash}m{.19\linewidth}|} \hline 
			Does isomorphism of designs coincide with EA-equivalence for $(n,m)$-bent functions? & Translation Designs     $\dev(\graph{F})$                            & Addition Designs      $\mathbb{D}(F)$                 \\ \hline
			$m=1$                                                                     & No, isomorphism is coarser for all $n\ge6$. & \multirow{2}{*}{Yes, for all $n$.} \\ \cline{1-2}
			$m\ge 2$                                                                     & Yes, for $n=4,6$.                                      &                                        \\ \hline
	\end{tabular}}
\end{table}
Finally, we would like to mention some open problems on bent functions and their translation designs, which the reader is invited to attack.
\begin{openproblem}
	As one can see from Examples~\ref{example: Inequivalent but isomorphic bents} and~\ref{example: Inequivalent but isomorphic bents 2}, it is possible to construct EA-inequivalent but isomorphic Boolean bent functions, by taking proper Maiorana-McFarland bent functions and extending them to infinite families using the Proposition~\ref{proposition: Isomorphisms of functions and their extensions}. So far, this approach does not seem to work for vectorial bent functions:
	\begin{itemize}
		\item there is only one up to EA-equivalence vectorial bent function in $4$ variables, from what follows that all derived translation designs are isomorphic;
		\item by Theorem~\ref{theorem: Isomorphic functions are equivalent} all isomorphic vectorial bent functions in 6 variables are also EA-equivalent. 
	\end{itemize}
	As one can see from Proposition~\ref{proposition: Isomorphisms of functions and their extensions}, a single example of EA-inequivalent but isomorphic vectorial bent functions will lead to an infinite family and, consequently, will prove that for vectorial bent functions the isomorphism of translation designs is a coarser equivalence relation than EA-equivalence. However, since one still does not have an example of such functions, it is essential to ask, whether EA-inequivalent but isomorphic vectorial bent functions may in general exist.
\end{openproblem}
\begin{openproblem}
	There are very few symmetric designs with a 2-transitive automorphism group, as it was shown by Kantor in~\cite{Kantor1985}. One of them is $2$-$(2^{n},2^{n-1}- 2^{n/2-1},2^{n-2} - 2^{n/2-1})$ design, which can be constructed as the addition $\D(f)$ or as the translation $\dev(D_f)$ design of a bent function $f$ on $\F_2^n$. While in the case of addition designs $\D(f)$ a bent function $f$ on $\F_2^n$ has to be quadratic, from Example~\ref{example: Inequivalent but isomorphic bents} one can see that for the translation design $\dev(\support{f})$ one still has some freedom to choose a function $f$. We conjecture, that a translation design $\dev(\support{f})$ of a bent function $f$ on $\F_2^n$ has 2-transitive automorphism group if and only if function $f$ is EA-equivalent to a Maiorana-McFarland bent function of the form $\langle \xBold, \yBold\rangle_{n/2}\oplus g(\yBold)$ with $\deg(g)\le 3$.
\end{openproblem}
\section*{Acknowledgment}
The authors would like to thank anonymous referees for their comments that helped to improve the presentation of the results.

\appendix
\section{Appendix: Algebraic normal forms of bent functions in six variables}
\label{appendix: A}

For any equivalence class $C^m_i$ of $(6,m)$-bent functions we compute the following invariants:
\begin{itemize}
	\item $|\Aut(C^{m}_{i})|:=|\Aut(\mathcal{C}(F^{m}_{i}))|$, that is the order of the automorphism group of the linear code $\mathcal{C}(F^{m}_{i})$, for a representative $F^{m}_{i}\in C^{m}_{i}$ listed in Appendix~\ref{appendix: A};
	\item $|\Aut(\dev(C^{m}_{i}))|:=|\Aut(\dev(\graph{F^{m}_{i}}))|$, that is the order of the automorphism group of the translation design $\dev(\graph{F^{m}_{i}})$, for a representative $F^{m}_{i}\in C^{m}_{i}$ listed in Appendix~\ref{appendix: A};
	\item $\snf(C^{m}_{i}):=\snf(F^{m}_{i})$, that is the \emph{Smith normal form} of the incidence matrix $M(\dev(\graph{F^{m}_{i}}))$, given by the multiset $\snf(F^{m}_{i})=\{*d_1^{e_1},\dots, d_{k}^{e_k}*\}$, where consecutive elementary divisors $d_k$ and $d_{k+1}$ satisfy $d_k|d_{k+1}$, and $e_k$ is the multiplicity of the elementary divisor $d_k$.
\end{itemize}

\noindent Note that the multiplicity of one in the Smith normal form $\snf(F^{m}_{i})$ is the $\Grank(F^{m}_{i})$ and similarly to~\cite[Proposition 2.4]{PolujanPott19DCC} one can show, that all elementary divisors $d_k$ in the $\snf(F)$ of an $(n,m)$-bent function $F$ are powers of two. We also observe that any two different equivalence classes $C^m_i$ and $C^m_j$ of bent functions in six variables have different pairs of invariants 
\begin{equation*}
(|\Aut(C^m_i)|,\snf(C^m_i)) \neq (|\Aut(C^m_j)|,\snf(C^m_j)).
\end{equation*}
\noindent In this way, the reader can be sure that all the representatives of equivalence classes listed in Appendix~\ref{appendix: A} are extended-affine inequivalent.

\begin{table}[H]
	\caption{Invariants of inequivalent $(6,m)$-bent functions}
	\label{table: Invariants of (6,m)-bent functions}
	\begin{subtable}{1\linewidth}
		\caption{Boolean $(6,1)$-bent functions}
		\label{table: Invariants of (6,1)-bent functions}
		\centering
		\scalebox{0.84}{\begin{tabular}{|c|c|c|c|}
				\hline
				$C^{1}_{i}$ & $|\Aut(C^{1}_{i})|$ & $\dfrac{|\Aut(\dev(C^{1}_{i}))|}{|\Aut(C^{1}_{i})|}$ &$\snf(C^{1}_{i})$  \\ \hline
				$C^{1}_{1}$ & $2^{15} \cdot 3^4 \cdot 5^1 \cdot 7^1$ & $2^ {13}\cdot 7^1 \cdot 31^1$ & $\left\{* 1^8,2^{15},4^{20},8^{15},16^6,32^1 *\right\}$ \\
				$C^{1}_{2}$ & $2^{15} \cdot 3^1 \cdot 7^1$ & $2^ {13}\cdot 3^3 \cdot 5^1 \cdot 7^1 \cdot 31^1$  &$\left\{* 1^8,2^{15},4^{20},8^{15},16^6,32^1 *\right\}$ \\
				$C^{1}_{3}$ & $2^{13} \cdot 3^1 \cdot 5^1$ & $2^{11} \cdot 3^1$ &$\left\{* 1^{12},2^9,4^{24},8^9,16^{10},32^1 *\right\}$ \\
				$C^{1}_{4}$ & $2^{11} \cdot 3^1 \cdot 7^1$ & $2^7$ &$\left\{* 1^{14},2^7,4^{24},8^7,16^{12},32^1 *\right\}$ \\ \hline
		\end{tabular}}
		\vspace{0.25cm}
	\end{subtable}
	
	\begin{subtable}{1\linewidth}
		\caption{Vectorial $(6,2)$-bent functions}
		\label{table: Invariants of (6,2)-bent functions}
		\centering
		\scalebox{0.8}{\begin{tabular}{|c|c|c|}
				\hline
				$C^{2}_{i}$ & $|\Aut(C^{2}_{i})|$ & $\snf(C^{2}_{i})$  \\ \hline
				$C^{2}_{1}$ & $2^9 \cdot  3^3 \cdot 7^1$  & $\left\{*1^{28},2^{26},4^{42},8^{64},16^{19},32^{12},64^2*\right\}$ \\ 
				$C^{2}_{2}$ & $2^9 \cdot  7^1$            & $\left\{*1^{30},2^{28},4^{40},8^{54},16^{27},32^{12},64^2*\right\}$ \\
				$C^{2}_{3}$ & $2^7 \cdot  3^1$            & $\left\{*1^{36},2^{22},4^{39},8^{50},16^{32},32^{12},64^2*\right\}$ \\
				$C^{2}_{4}$ & $2^6$                       & $\left\{*1^{38},2^{24},4^{33},8^{56},16^{20},32^{20},64^2*\right\}$ \\
				$C^{2}_{5}$ & $2^6 \cdot 3^1$             & $\left\{*1^{38},2^{24},4^{37},8^{48},16^{24},32^{20},64^2*\right\}$ \\
				$C^{2}_{6}$ & $2^4$                       & $\left\{*1^{42},2^{20},4^{37},8^{48},16^{20},32^{24},64^2*\right\}$ \\
				$C^{2}_{7}$ & $2^4 \cdot  3^1 \cdot  7^1$ & $\left\{*1^{36},2^{34},4^{23},8^{58},16^{16},32^{24},64^2*\right\}$ \\
				$C^{2}_{8}$ & $2^1 \cdot  7^1$            & $\left\{*1^{42},2^{22},4^{41},8^{34},16^{28},32^{24},64^2*\right\}$ \\
				$C^{2}_{9}$ & $2^1 \cdot 3^1 \cdot  7^1$  & $\left\{*1^{42},2^{22},4^{41},8^{34},16^{28},32^{24},64^2*\right\}$ \\ \hline
		\end{tabular}}
		\vspace{0.25cm}	
	\end{subtable}%
	
	\begin{subtable}{1\linewidth}
		\caption{Vectorial $(6,3)$-bent functions}
		\label{table: Invariants of (6,3)-bent functions}
		\centering
		\scalebox{0.8}{\begin{tabular}{|c|c|c|}
				\hline
				$C^{3}_{i\phantom{0}}$  & $|\Aut(C^{3}_{i})|$       & $\snf(C^{3}_{i})$  \\ \hline
				$C^{3}_{1\phantom{0}}$  & $2^9 \cdot 3^3 \cdot 7^2$ & $\left\{*1^{64\phantom{0}},2^{48},4^{72},8^{163},16^{54},32^{30},64^{18}*\right\}$ \\
				$C^{3}_{2\phantom{0}}$  & $2^9\cdot3^1\cdot7^1$  	 & $\left\{*1^{78\phantom{0}},2^{44},4^{68},8^{139},16^{62},32^{38},64^{20}*\right\}$ \\
				$C^{3}_{3\phantom{0}}$  & $2^6\cdot3^2\cdot7^1$  	 & $\left\{*1^{88\phantom{0}},2^{32},4^{68},8^{137},16^{68},32^{32},64^{24}*\right\}$ \\
				$C^{3}_{4\phantom{0}}$  & $2^6\cdot3^1\cdot7^1$  	 & $\left\{*1^{80\phantom{0}},2^{40},4^{70},8^{145},16^{54},32^{36},64^{24}*\right\}$ \\
				$C^{3}_{5\phantom{0}}$  & $2^3\cdot3^1\cdot7^1$  	 & $\left\{*1^{88\phantom{0}},2^{48},4^{48},8^{145},16^{48},32^{48},64^{24}*\right\}$ \\
				$C^{3}_{6\phantom{0}}$  & $2^4\cdot3^1$ 		  	 & $\left\{*1^{98\phantom{0}},2^{44},4^{38},8^{153},16^{38},32^{44},64^{34}*\right\}$ \\
				$C^{3}_{7\phantom{0}}$  & $2^3\cdot3^1$ 		  	 & $\left\{*1^{98\phantom{0}},2^{40},4^{46},8^{145},16^{46},32^{40},64^{34}*\right\}$ \\
				$C^{3}_{8\phantom{0}}$  & $2^2\cdot3^1\cdot7^1$  	 & $\left\{*1^{100},2^{36},4^{36},8^{169},16^{36},32^{36},64^{36}*\right\}$ \\
				$C^{3}_{9\phantom{0}}$  & $2^2\cdot3^1 $         	 & $\left\{*1^{106},2^{36},4^{30},8^{169},16^{30},32^{36},64^{42}*\right\}$ \\
				$C^{3}_{10}$ & $2^3\cdot3^1\cdot7^1 $ 	 & $\left\{*1^{100},2^{36},4^{36},8^{169},16^{36},32^{36},64^{36}*\right\}$ \\
				$C^{3}_{11}$ & $2^3\cdot3^1\cdot7^2$  	 & $\left\{*1^{100},2^{36},4^{36},8^{169},16^{36},32^{36},64^{36}*\right\}$ \\
				$C^{3}_{12}$ & $3^1\cdot7^1 $         	 & $\left\{*1^{106},2^{42},4^{18},8^{181},16^{18},32^{42},64^{42}*\right\}$ \\
				$C^{3}_{13}$ & $3^1\cdot7^1 $         	 & $\left\{*1^{106},2^{36},4^{30},8^{169},16^{30},32^{36},64^{42}*\right\}$ \\ \hline
		\end{tabular}}
	\end{subtable}	
\end{table}
\noindent For quadratic vectorial bent functions from equivalence classes $C^m_1$ with $m=2,3$ we have $|\Aut(\dev(C^{m}_{1}))| =2^{n+m} \cdot |\Aut(C^{m}_{1})|\cdot 7^1$, where $n=6$. For the rest of vectorial $(n,m)$-bent functions in six variables we have $|\Aut(\dev(C^{m}_{i}))| =2^{n+m} \cdot |\Aut(C^{m}_{i})|$. In this way, translation designs of vectorial bent functions from equivalence classes $C^3_{10}$ and $C^3_{11}$ can be distinguished by the orders of their automorphism groups, despite the Smith normal forms coincide.

Finally, we list algebraic normal forms of representatives of EA-equivalence classes of bent functions in $6$ variables. The representatives $F^{m}_i\in C^{m}_i$ and $F^{m+1}_j\in C^{m+1}_j$ are selected in such a way, that $F^{m}_i \prec F^{m+1}_j$ as on Figure~\ref{figure: Hasse Diagram}. We abbreviate $1\le i\le6$ for the variable $x_i$. 

\begin{table*}
	\caption{Algebraic normal forms of inequivalent $(6,m)$-bent functions}
	\label{table: Algebraic normal forms of inequivalent (6,m)-bent functions}
	\begin{subtable}{1\linewidth}
		\caption{Boolean $(6,1)$-bent functions}
		\label{table: ANFs of (6,1)-bent functions}
		\centering
		\scalebox{1}{
			\begin{tabular}{|c|c|}
				\hline
				$F^{1}_{i}$ & Algebraic normal form of $F^{1}_{i}\in C^{1}_{i}$ \\ \hline
				$F^{1}_{1}$ & $1 4 \oplus  2 5 \oplus 3 6 $  \\ 
				$F^{1}_{2}$ & $1 4 \oplus  2 5 \oplus 3 6 \oplus 1 2 3$\\
				$F^{1}_{3}$ & $1 2 \oplus  1 4 \oplus 2 6 \oplus  3 5 \oplus  4 5 \oplus 1 2 3 \oplus 2 4 5$ \\
				$F^{1}_{4}$ & $1 4 \oplus  2 6 \oplus 3 4 \oplus  3 5 \oplus  3 6 \oplus  4 5 \oplus  4 6 \oplus  1 2 3 \oplus  2 4 5  \oplus  3 4 6 $ \\ \hline
			\end{tabular}
		}
		\vspace{0.2cm}
	\end{subtable}
	
	\begin{subtable}{1\linewidth}
		\caption{Vectorial $(6,2)$-bent functions}
		\label{table: ANFs of (6,2)-bent functions}
		\centering
		\scalebox{1}{
			\begin{tabular}{|c|c|}
				\hline
				$F^{2}_{i}$ & Algebraic normal form of $F^{2}_{i}\in C^{2}_{i}$  \\ \hline
				$F^{2}_{1}$ & $ \begin{pmatrix}
				1 4 \oplus  2 5 \oplus 3 6 \\
				1 5 \oplus 1 6 \oplus 2 4 \oplus 2 5 \oplus 3 4
				\end{pmatrix}$  \\ 
				$F^{2}_{2}$ & $ \begin{pmatrix}
				1 4 \oplus  2 5 \oplus 3 6 \oplus 1 2 3\\
				1 5 \oplus 1 6 \oplus 2 4 \oplus 2 5 \oplus 3 4
				\end{pmatrix}$\\
				$F^{2}_{3}$ & $\begin{pmatrix} 
				1 4 \oplus  2 5 \oplus 3 6 \oplus 1 2 3 \\
				1 3 \oplus  1 5 \oplus  2 3 \oplus  4 6 \oplus 1 2 4 \end{pmatrix}$ \\
				$F^{2}_{4}$ & $\begin{pmatrix}
				1 4 \oplus  2 5 \oplus 3 6 \oplus 1 2 3 \\
				1 2 \oplus  1 3 \oplus 1 6 \oplus  2 6 \oplus  4 5 \oplus  5 6 \oplus  1 5 6 \oplus 2 3 5 \end{pmatrix}$ \\
				$F^{2}_{5}$ & $\begin{pmatrix}
				1 2 \oplus  1 4 \oplus 2 6 \oplus  3 5 \oplus  4 5 \oplus 1 2 3 \oplus 2 4 5 \\
				1 3 \oplus  2 3 \oplus  2 4 \oplus  3 5 \oplus  5 6 \oplus 1 2 6 \oplus  2 3 5 \end{pmatrix}$ \\
				$F^{2}_{6}$ & $\begin{pmatrix}
				1 4 \oplus  2 6 \oplus 3 4 \oplus  3 5 \oplus 3 6 \oplus 4 5 \oplus 4 6 \oplus 1 2 3 \oplus 2 4 5 \oplus 3 4 6 \\ 
				1 3  \oplus  2 3 \oplus  2 4 \oplus  3 5 \oplus  5 6 \oplus 1 2 6 \oplus  2 3 5	\end{pmatrix}$ \\
				$F^{2}_{7}$ & $\begin{pmatrix}
				1 4 \oplus 2 6 \oplus 3 4 \oplus 3 5 \oplus 3 6 \oplus 4 5 \oplus 4 6 \oplus 1 2 3 \oplus 2 4 5 \oplus 3 4 6 \\
				1 2 \oplus 3 5 \oplus  4 6 \oplus 1 2 4 \oplus  1 3 4 \oplus  2 3 5 \oplus  2 3 6 \oplus  2 4 5 \end{pmatrix}$ \\
				$F^{2}_{8}$ & $\begin{pmatrix}
				1 4 \oplus 2 6 \oplus 3 4 \oplus 3 5 \oplus 3 6 \oplus 4 5 \oplus 4 6 \oplus 1 2 3 \oplus 2 4 5 \oplus  3 4 6 \\
				1 2 \oplus  1 6 \oplus  2 3  \oplus  3 5 \oplus  4 6 \oplus  5 6 \oplus 1 2 4 \oplus  1 3 4 \oplus  1 5 6  \oplus  2 3 5 \oplus  2 3 6 \oplus  2 4 5 \end{pmatrix}$ \\
				$F^{2}_{9}$ & $\begin{pmatrix}
				1 4 \oplus 2 6 \oplus 3 4 \oplus 3 5 \oplus 3 6 \oplus 4 5 \oplus 4 6 \oplus 1 2 3 \oplus 2 4 5 \oplus 3 4 6 \\
				1 2 \oplus  1 5 \oplus  1 6 \oplus  2 5 \oplus  3 6 \oplus  4 5 \oplus  4 6 \oplus 1 2 5 \oplus  1 2 6 \oplus  1 3 5 \oplus  1 3 6 \oplus  1 4 5 \oplus  2 5 6 \end{pmatrix}$  \\ \hline
			\end{tabular}
		}
		\vspace{0.2cm}	
	\end{subtable}%
	
	\begin{subtable}{1\linewidth}
		\caption{Vectorial $(6,3)$-bent functions}
		\label{table: ANFs of (6,3)-bent functions}
		\centering
		\scalebox{1}{
			\begin{tabular}{|c|c|}
				\hline
				$F^{3}_{i}$ & Algebraic normal form of $F^{3}_{i}\in C^{3}_{i}$  \\ \hline
				$F^{3}_{1\phantom{0}}$ & $ \begin{pmatrix}
				1 4 \oplus  2 5 \oplus 3 6 \\
				1 5 \oplus 1 6 \oplus 2 4 \oplus 2 5 \oplus 3 4 \\
				1 4 \oplus  1 5 \oplus  2 4 \oplus  2 5 \oplus  2 6 \oplus  3 5
				\end{pmatrix}$  \\ 
				$F^{3}_{2\phantom{0}}$ & $ \begin{pmatrix}
				1 4 \oplus  2 5 \oplus 3 6 \\
				1 5 \oplus 1 6 \oplus 2 4 \oplus 2 5 \oplus 3 4\\
				1 2 \oplus  1 4 \oplus  1 5 \oplus  2 4 \oplus  2 5 \oplus  2 6 \oplus  3 5
				\end{pmatrix}$  \\
				$F^{3}_{3\phantom{0}}$ & $ \begin{pmatrix}
				1 4 \oplus  2 5 \oplus 3 6 \\
				1 5 \oplus 1 6 \oplus 2 4 \oplus 2 5 \oplus 3 4\\
				1 3 \oplus  1 4 \oplus  2 6 \oplus  4 5
				\end{pmatrix}$  \\ 
				$F^{3}_{4\phantom{0}}$ & $ \begin{pmatrix}
				1 4 \oplus  2 5 \oplus 3 6 \\
				1 5 \oplus 1 6 \oplus 2 4 \oplus 2 5 \oplus 3 4 \\
				1 4 \oplus  1 5 \oplus  2 4 \oplus  2 5 \oplus  2 6 \oplus  3 5 \oplus 1 2 3
				\end{pmatrix}$  \\ 
				$F^{3}_{5\phantom{0}}$ & $ \begin{pmatrix}
				1 4 \oplus  2 5 \oplus 3 6 \oplus 1 2 3 \\
				1 3 \oplus  1 5 \oplus  2 3 \oplus  4 6 \oplus 1 2 4\\
				1 3 \oplus  2 4 \oplus  2 5 \oplus  5 6 \oplus 1 2 5
				\end{pmatrix}$  \\ 
				$F^{3}_{6\phantom{0}}$ & $ \begin{pmatrix}
				1 4 \oplus  2 5 \oplus 3 6 \oplus 1 2 3 \\
				1 3 \oplus  1 5 \oplus  2 3 \oplus  4 6 \oplus 1 2 4 \\
				1 2 \oplus  1 4 \oplus  1 6 \oplus 3 4 \oplus  4 6 \oplus  5 6 \oplus  1 2 6 \oplus  1 3 6 \oplus  2 4 6
				\end{pmatrix}$  \\ 
				$F^{3}_{7\phantom{0}}$ & $ \begin{pmatrix} 
				1 4 \oplus  2 5 \oplus 3 6 \oplus 1 2 3 \\
				1 3 \oplus  1 5 \oplus  2 3 \oplus  4 6 \oplus 1 2 4\\
				1 2 \oplus  1 3 \oplus  2 4 \oplus  2 5 \oplus  3 5 \oplus  4 5 \oplus  5 6 \oplus 1 2 5 \oplus  3 4 5
				\end{pmatrix}$  \\
				$F^{3}_{8\phantom{0}}$ & $ \begin{pmatrix}
				1 2 \oplus  1 4 \oplus 2 6 \oplus  3 5 \oplus  4 5 \oplus 1 2 3  \oplus 2 4 5 \\
				1 3 \oplus  2 3 \oplus  2 4 \oplus  3 5 \oplus  5 6 \oplus 1 2 6 \oplus  2 3 5 \\
				1 6 \oplus  2 3 \oplus  2 6 \oplus  3 5 \oplus  4 5 \oplus  5 6	 \oplus 1 2 3 \oplus  1 2 4 \oplus  2 5 6
				\end{pmatrix}$  \\ 
				$F^{3}_{9\phantom{0}}$ & $ \begin{pmatrix}
				1 2 \oplus  1 4 \oplus 2 6 \oplus  3 5 \oplus  4 5 \oplus 1 2 3 \oplus 2 4 5 \\
				1 3 \oplus  2 3 \oplus  2 4 \oplus  3 5 \oplus  5 6 \oplus 1 2 6 \oplus  2 3 5\\ 
				1 6 \oplus  2 5 \oplus  2 6 \oplus  3 5 \oplus  3 6 \oplus  4 5 \oplus  5 6 \oplus 1 2 3 \oplus  1 2 4 \oplus  2 3 4 \oplus  2 5 6 \oplus  3 4 6
				\end{pmatrix}$  \\
				$F^{3}_{10}$ & $ \begin{pmatrix}
				1 4 \oplus 2 6  \oplus 3 4  \oplus 3 5 \oplus 3 6 \oplus 4 5 \oplus 4 6 \oplus 1 2 3 \oplus 2 4 5 \oplus 3 4 6 \\
				1 2 \oplus 3 5  \oplus  4 6 \oplus 1 2 4 \oplus  1 3 4 \oplus  2 3 5 \oplus  2 3 6 \oplus  2 4 5\\ 
				1 2 \oplus  1 3 \oplus  2 5 \oplus  3 5 \oplus  3 6 \oplus  4 5	\oplus 1 2 3 \oplus  1 3 4 \oplus  2 3 6 \oplus  2 4 6 \oplus  3 4 5
				\end{pmatrix}$  \\ 
				$F^{3}_{11}$ & $ \begin{pmatrix}
				1 4 \oplus 2 6  \oplus 3 4 \oplus 3 5 \oplus 3 6 \oplus 4 5 \oplus 4 6 \oplus 1 2 3 \oplus 2 4 5 \oplus 3 4 6 \\
				1 2 \oplus 3 5  \oplus  4 6 \oplus 1 2 4 \oplus  1 3 4 \oplus  2 3 5 \oplus  2 3 6 \oplus  2 4 5\\
				1 2 \oplus  1 3 \oplus  2 4 \oplus  2 5  \oplus  3 4 \oplus  3 5 \oplus  3 6  \oplus  4 5 \oplus 1 2 3 \oplus  1 3 4 \oplus  2 3 6 \oplus  2 4 6 \oplus  3 4 5
				\end{pmatrix}$  \\ 
				$F^{3}_{12}$ & $ \begin{pmatrix}
				1 4 \oplus 2 6 \oplus 3 4   \oplus 3 5 \oplus 3 6 \oplus 4 5 \oplus 4 6 \oplus 1 2 3 \oplus 2 4 5 \oplus 3 4 6 \\
				1 2 \oplus 3 5 \oplus  4 6  \oplus 1 2 4 \oplus  1 3 4 \oplus  2 3 5 \oplus  2 3 6 \oplus  2 4 5\\
				1 4 \oplus  1 5 \oplus  1 6 \oplus  2 3  \oplus  2 6 \oplus  3 5 \oplus  5 6 \oplus 1 2 4 \oplus  1 2 5 \oplus  1 2 6 \oplus  1 3 6 \oplus  1 4 5 \oplus  1 5 6 \oplus  2 3 6  \oplus  2 4 6 \oplus  3 4 5
				\end{pmatrix}$  \\ 
				$F^{3}_{13}$ & $ \begin{pmatrix}
				1 4 \oplus 2 6 \oplus 3 4 \oplus 3 5 \oplus 3 6 \oplus 4 5 \oplus 4 6 \oplus 1 2 3 \oplus 2 4 5 \oplus 3 4 6 \\
				1 2 \oplus 3 5 \oplus  4 6 \oplus 1 2 4 \oplus  1 3 4 \oplus  2 3 5 \oplus  2 3 6 \oplus  2 4 5\\
				1 3 \oplus  1 4 \oplus  2 4 \oplus  3 4  \oplus  3 5 \oplus  3 6 \oplus  4 6 \oplus  5 6 \oplus 1 2 3 \oplus  1 2 5 \oplus  1 4 5 \oplus  1 4 6 \oplus  2 3 5 \oplus  2 5 6\oplus  3 5 6 \oplus  4 5 6
				\end{pmatrix}$  \\ \hline
			\end{tabular}
		}
	\end{subtable}	
\end{table*}
\clearpage{\phantomsection}
\begin{IEEEbiographynophoto}{Alexandr A. Polujan}
	received Diploma in mathematics and system analysis from the Faculty of Mechanics and Mathematics of the Belarusian State University, Minsk, Belarus, in 2015. Since 2016, he has been pursuing the Ph.D. degree under the supervision of Alexander Pott at the Otto von Guericke University, Magdeburg, Germany. His research interests include mainly perfect nonlinear functions.
\end{IEEEbiographynophoto}
\vskip 0pt plus -1fil
\begin{IEEEbiographynophoto}{Alexander Pott}
	received his Ph.D. degree in Mathematics in 1988 from the Justus-Liebig-University Giessen. He held visiting positions in Dayton (Ohio) and Duisburg (Germany). He was professor at the University of Augsburg (Germany), and presently he is full professor at the Otto von Guericke University Magdeburg, Germany. His research interests include Finite Geometry (difference sets), Algebraic Coding Theory (sequences, Boolean functions) and Finite Fields.
\end{IEEEbiographynophoto}
\end{document}